\documentclass[iop,12pt]{emulateapj}
\pdfoutput=1
\usepackage{apjfonts}
\usepackage{epsfig}
\usepackage{natbib}
\usepackage{amsfonts}
\usepackage{amsmath}
\usepackage{booktabs}
\usepackage{multirow}
\usepackage{enumerate}
\usepackage[usenames]{color}
\newcommand*{\sym}{\mathord{\sim}}

\def\Dwa{$\,$\uppercase\expandafter{\romannumeral5}$\,$}

\def\sless{\lower2pt\hbox{$\buildrel {\scriptstyle <}
   \over {\scriptstyle\sim}$}}

\def\sgreat{\lower2pt\hbox{$\buildrel {\scriptstyle >}
   \over {\scriptstyle\sim}$}}
\def\sharpnull#1{}

\newcommand{\Ye}{Y_e}
\newcommand{\avgE}{\langle E_{\nu_i} \rangle}
\newcommand{\Rnui}{R_{\nu_i}}

\newcommand{\code}[1]{\texttt{#1}}

\begin{document}
\slugcomment{Published in ApJ.}

\title{
Black hole--neutron star mergers with a hot nuclear equation of state:
\\ outflow and neutrino-cooled disk for a low-mass, high-spin case
}

\author{M. Brett Deaton\altaffilmark{1}}
\author{Matthew D. Duez\altaffilmark{1}}
\author{Francois Foucart\altaffilmark{2}}
\author{Evan O'Connor\altaffilmark{2}}
\author{Christian D. Ott\altaffilmark{4,5}}
\author{Lawrence~E.~Kidder\altaffilmark{3}}
\author{Curran D. Muhlberger\altaffilmark{3}}
\author{Mark A. Scheel\altaffilmark{4}}
\author{Bela Szilagyi\altaffilmark{4}}
  \altaffiltext{1}{Department of Physics \& Astronomy,
    Washington State University, Pullman, Washington 99164, USA;
    mbdeaton@wsu.edu, m.duez@wsu.edu}
  \altaffiltext{2}{Canadian Institute for Theoretical
    Astrophysics, University of Toronto, Toronto, Ontario M5S 3H8, Canada}
  \altaffiltext{3}{Center for Radiophysics and Space
    Research, Cornell University, Ithaca, New York, 14853, USA}
  \altaffiltext{4}{TAPIR, MC 350-17,
    California Institute of Technology, Pasadena, California 91125, USA}
  \altaffiltext{5}{Alfred P. Sloan Research Fellow}

\begin{abstract}
  Neutrino emission significantly affects the evolution of the
  accretion tori formed in black hole--neutron star mergers.  It
  removes energy from the disk, alters its composition, and provides a
  potential power source for a gamma-ray burst.  To study these
  effects, simulations in general relativity with a hot microphysical
  equation of state and neutrino feedback are needed.  We present the
  first such simulation, using a neutrino leakage scheme for cooling
  to capture the most essential effects and considering a moderate
  mass (1.4~$M_{\odot}$ neutron star, 5.6~$M_{\odot}$ black hole),
  high spin (black hole $J/M^2=0.9$) system with the $K_0=220$~MeV
  Lattimer-Swesty equation of state.  We find that about 0.08~$M_{\odot}$ of
  nuclear matter is ejected from the system, while another
  0.3~$M_{\odot}$ forms a hot, compact accretion disk.  The primary
  effects of the escaping neutrinos are (\emph{i}) to make the disk
  much denser and more compact, (\emph{ii}) to cause the average
  electron fraction $Y_e$ of the disk to rise to about 0.2 and then
  gradually decrease again, and (\emph{iii}) to gradually cool the
  disk.  The disk is initially hot ($T \sim 6$~MeV) and luminous
  in neutrinos ($L_{\nu}\sim 10^{54}$~erg~s${}^{-1}$), but the
  neutrino luminosity decreases by an order of magnitude over 50~ms
  of post-merger evolution.
\end{abstract}

\keywords{Black hole physics -- Gamma-ray burst: general -- Neutrinos -- Stars: neutron}

\section{Introduction}

Much of the interest in black hole--neutron star (BHNS) mergers
arises from their potential to solve two important problems in contemporary
astrophysics.  First, it is possible that such events can produce
short hard gamma-ray bursts.  Second, they could significantly
contribute to the abundances of r-process nuclei observed in the solar
system.

Short hard gamma-ray bursts (SGRBs) emit characteristic luminosities of
$L \sim 10^{50\text{--}52}/(4\pi)\rm\,erg\,s^{-1}\,steradian^{-1}$,
with typical durations of $\sym 1\rm\,s$,
and peak photon energies of $h\nu \sim 1$~MeV.
Their spectra are nonthermal,
and vary in brightness on a timescale of $\Delta t \lesssim 10$~ms \citep{Nakar2007}.
Combined observations of a nonthermal spectrum and a short
time variability can be explained by an ultra-relativistic jet
($\Gamma \gtrsim 30$) launched from a disk experiencing rapid accretion onto
a black hole \citep{1992ApJ...395L..83N,Nakar2007}.
BHNS mergers are plausible progenitors for such a system.

BHNS mergers may also produce unbound neutron-rich ejecta, which
would provide an r-process nucleosynthesis site needed to explain
the observed heavy element abundances
\citep{1974ApJ...192L.145L,Freiburghaus:1999no,Arnould:2007gh,korobkin:12}.
Additionally such outflows may be observable from the radioactive decay
that results after unstable heavy isotopes are formed (a `kilonova')
or from the shock that would form when the outflow hits a sufficiently
dense interstellar medium \citep{metzger:11}.

BHNS mergers and their aftermath involve relativistic gravity,
magnetohydrodynamics, nuclear physics, and neutrino radiation. 
In addition, a large space of binary parameters must be explored,
since a wide range of premerger black hole mass and spin values
are plausible \citep{belczynski:08,2010ApJ...725.1918O}.
Early attempts to explore this parameter space
in general relativity have used polytropic \citep{Taniguchi:2005fr,
Shibata:2006ks,Etienne:2007jg,Etienne:2008re,Duez:2008rb,Duez:2009yy,
Foucart:2010eq,FoucartEtAl:2011,Etienne:2012te,Foucart:2013a} or
piecewise-polytropic \citep{Kyutoku:2010zd,Kyutoku:2011vz,Lackey:2013}
equations of state (EOS) to describe the neutron star matter, with the
thermal effects being modeled by a simple $\Gamma$-law. Because
the matter does not heat until after tidal disruption,
use of these $\Gamma$-law EOSs may be adequate for the inspiral and very early
tidal disruption phases, which produce nearly all of the gravitational wave signal.
From these parameter space studies, a general picture has begun to emerge
\citep{faber:2009review,Duez:2009yz,Pannarale:2010vs,shibata:2011,Foucart2012}.
In the best-understood binary parameter space of
spin ($a^{*} \equiv (J_{\rm BH}/M_{\rm BH}{}^2)(c/G) = [-0.5,0.9]$),
NS compactness ($\mathcal{C} \equiv (M_{\rm NS}/R_{\rm NS})(G/c^2) \sim [0.13,0.20]$),
and mass ratio ($q \equiv M_{\rm BH}/M_{\rm NS} \sim [1,7]$)---where
G is Newton's gravitational constant and c is the speed of light---disks
of significant masses appear to be formed in binaries with low mass ratios,
low NS compactness, or high prograde black hole spins.

After the merger, neutrinos will begin to play an important
role in the evolution. Following disruption, the fluid
is shocked to temperatures of $\sym10$~MeV and begins to radiate copiously.
In simulations of accretion disk--black hole systems
\citep{Shibata:2012zz},
and binary neutron star mergers \citep{1997A&A...319..122R,Rosswog:2003rv,Sekiguchi:2011zd},
which produce disks similar to those of BHNS mergers,
total neutrino luminosities are of order
$10^{51\text{--}53}\rm\,erg\,s^{-1}$.
BHNS simulations using $\Gamma$-law EOSs
indicate that disks last at most several hundred
milliseconds \citep{Etienne:2008re,Duez:2009yy,Kyutoku:2010zd,
Foucart:2010eq}.  During this time, neutrinos may carry away
a significant amount of energy from the accretion disk, causing
it to cool and contract.  Additionally, neutrinos, unlike photons,
change the composition of their source.
In the present case, initially neutron-rich neutron star material
releptonizes as it expands into a disk
through an imbalance of electron- and positron-capture reactions.
These composition changes can play a significant role in
BHNS post-merger dynamics in a number of ways.
Changes to the electron fraction alter the luminosity of each neutrino species.
Lepton number gradients affect pressure forces and can
drive convection \citep{Lee:2005se}.
Phase transitions, e.g.\ recombination from pure nucleonic matter
into alpha particles and heavy nuclei,
can release or absorb energy as density and temperature change.
A temperature- and composition-dependent
EOS is needed to capture all of these effects.

Neutrino emission may also play a crucial role in powering the
relativistic outflow needed for an SGRB.  It could do this by
depositing energy in the funnel region along the BH spin axis
via neutrino pair annihilation ($\nu\bar{\nu} \rightarrow e^{-}e^{+}$).
Depending on the shape of the emitting region, its proximity to the
black hole, and its emission spectrum, the annihilation process could
proceed with an efficiency as high as 0.2\%--0.5\%
\citep{1997A&A...319..122R,Birkl2007,dessart2009,Harikae:2010yt}.
This would indicate that neutrino pair annihilation
may only be able to power low-energy SGRBs.  However,
accurate relativistic simulations of the fluid and neutrino fields
are needed to make reliable statements about the efficiency of the
neutrino pair annihilation process in BHNS mergers.

Because the disk spans optically thick and optically thin regimes,
a full 6-dimensional evolution of the neutrino fields
(1 energetic + 3 spatial + 2 angular dimensions) is needed to
completely describe the coupling between radiation and matter.
This is impossible with current resources.
Fortunately, many of the essential features of the radiation can be
captured using a simple {\it leakage} scheme.  Rather than performing
actual radiation transport, leakage schemes remove energy and
alter lepton number at rates based on the local free-emission and
diffusion rates.  Leakage certainly neglects some arguably important effects (e.g.
neutrino absorption-driven winds), but it captures the basic energetics
and composition drift of the post-merger system.
(See e.g.\ \citealt{Ott:2012mr} for an analysis of leakage in core-collapse
supernovae simulations.)

In Newtonian gravity, microphysical EOSs and
neutrino leakage schemes have been used in simulations of binary neutron star
\citep{Ruffert1996,1997A&A...319..122R,2002MNRAS.334..481R,
Rosswog:2003rv} and BHNS \citep{Janka1999,2004MNRAS.351.1121R} mergers. 
Leakage schemes have recently been incorporated into three-dimensional general relativistic
simulations of stellar core collapse \citep{2011ApJ...737....6S,
Ott:2012mr} and binary neutron star mergers \citep{Sekiguchi:2011zd,Kiuchi:2012mk}. 
Simulations of BHNS mergers with a microphysical EOS but
no neutrino feedback were presented by \cite{Duez:2009yy}. 
The first relativistic BHNS simulations with both a microphysical EOS
and neutrino feedback are presented here.

In this paper, we introduce a neutrino leakage scheme into the
Spectral Einstein Code\footnote{\url{http://www.black-holes.org/SpEC.html}},
evolving the coupled hydrodynamics--Einstein field equations with leakage,
and apply it to one BHNS case, a moderately low-mass
($M_{\rm NS} = 1.4\,M_{\odot}$,
$M_{\rm BH} = 5.6\,M_{\odot}$), high black hole spin ($a^{*}=0.9$) system.
We follow six orbits of inspiral and evolve for 50~ms past merger.
We employ the finite-temperature EOS of
\cite{Lattimer:1991nc} using a compressibility parameter $K_0=220$~MeV
that yields neutron stars consistent with existing observations
(see e.g.\ \citealt{Demorest:2010bx} and \citealt{Steiner2010}---though recent work by
\citealt{Guillot:2013wu} predicts smaller neutron star radii).

We find an initially large postmerger disk of mass
$M_0 \sim 0.3\,M_{\odot}$ which accretes rapidly for the first
30\,ms and then settles to a slow mass depletion rate, from which we
can estimate a disk lifetime of $T_{\rm dep} \gtrsim 0.2$~s.
The neutrino luminosity
peaks 10~ms after merger at $\sym 10^{54}\rm\,erg\,s^{-1}$ but drops to 
$\sym 2 \times 10^{53}\rm\,erg\,s^{-1}$ by the end of the 50~ms post-merger
evolution, at which time it continues to drop.
Assuming a conservative efficiency of 0.2\% \citep{1997A&A...319..122R},
we estimate that the peak work done by neutrino pair annihilation
could be as high as $Q_{\nu\bar{\nu}} \sim 10^{51}\rm\,erg\,s^{-1}$.
We also find a large unbound outflow from the system, of mass
$M_{\rm ej} \sim 0.08\,M_{\odot}$, with mildly relativistic velocities,
leading to an available kinetic energy that could be as high as $E_{\rm ej} \sim 10^{52}$~erg.

The disk itself displays an interesting evolution during this time. 
A very high energy region---both hot and having super-Keplerian kinetic
energies---develops in the innermost disk.  This configuration is
unstable, and nonaxisymmetric perturbations persist in the inner
disk for many dynamical times.  Away from the edges, the disk does
roughly settle to stationary axisymmetry, and evolution is driven
primarily by secular radiation and disk depletion effects. 
Neutrino cooling leads to
a much higher-density, and somewhat lower-entropy, disk structure than seen in
a comparison run without cooling.  The average temperature
decreases in the disk from a maximum of $\sym 6$~MeV at 15~ms after merger
to $\sym 4$~MeV at 50~ms.
The average specific entropy ($\sym 0.1\,k_{\rm B}\,\rm baryon^{-1}$ during
inspiral) stays near $\sym 8\,k_{\rm B}\,\rm baryon^{-1}$
from 5~ms to the end of the evolution.
The disk is sufficiently
hot and dense to remain opaque to neutrinos of all species.
The average electron fraction rises from 0.07 in the initial neutron star
to around 0.2 at 20~ms after merger.
At this time, electron neutrino and antineutrino number emission
roughly balance.  Afterwards, the electron fraction
decreases gradually as the disk cools and the balance of
$\nu_e$ and $\bar{\nu}_e$ emission adjusts. The disk continues
to cool and flatten, so that by the end of our simulation
its luminosity in neutrinos has fallen by an order of magnitude.
Longer-lasting high-power energy release may come from
physics not included in this simulation, particularly magnetic fields.

This paper is organized as follows.
In Section~\ref{sec:Methods}, we discuss the numerical methods used to
simulate the mergers;
in subsections~\ref{sec:EOS}, \ref{sec:ID}, and \ref{sec:leakage},
we describe our treatment of the nuclear EOS, initial data, and neutrino
leakage, respectively.
In Section~\ref{sec:grhydroconvergence}, we give a summary of the
evolution, examining several measures of convergence.
In Section~\ref{sec:outflow}, we analyze outflows.
In Section~\ref{sec:scales}, we review the relevant timescales for the
accretion disk.
In Sections~\ref{sec:disk} and~\ref{sec:neutrinodisk}, we analyze the
accretion disk in the epochs of formation and neutrino-driven evolution,
respectively.
The importance of neutrino cooling effects in general is demonstrated in
Section~\ref{sec:comparison}, in which we compare the results of
simulations performed with and without neutrino leakage effects.
We summarize our conclusions in Section~\ref{sec:conclusion}.

Throughout the rest of this paper, unless otherwise noted, we adopt geometric units in which $G=c=1$,
and we use Latin indices $(j,k)$ to represent the three spatial coordinates,
and Greek indices $(\alpha,\beta)$ to represent the four spacetime coordinates.
Unless a different average is specified, angle brackets around thermodynamic
quantities indicate a density-weighted average
(e.g.\ $\langle Y_e \rangle \equiv \int Y_e \rho\, d^3x/ \int \rho\, d^3x$).

\section{Methods}
\label{sec:Methods}

We evolve the coupled general relativistic--hydrodyamics system using
the Spectral Einstein Code (SpEC).
We employ our standard two-grid pseudospectral/finite difference approach,
described in detail in earlier papers
\citep{Scheel2006,Duez:2008rb,Hemberger:2012jz,Foucart:2013a}
and briefly summarized here. The spacetime is described by the 4-metric and
its time derivative, which we evolve in the generalized harmonic formulation using a multidomain
pseudospectral algorithm \citep{Lindblom:2007}.
The coordinates are evolved by enforcing the `frozen' gauge condition described
in Appendix~A.1 of \cite{Foucart:2013a}.
The fluid is described by the hydrodynamic fields,
which we evolve in conservative form using shock-capturing finite-difference
methods.  The metric and fluid are evolved on separate computational domains;
fluid source terms for the metric evolution and metric source terms for the
fluid evolution are acquired by interpolation between domains.  The fluid
domain is a uniform Cartesian grid covering the non-vacuum upper hemisphere,
with the lower hemisphere fluid quantities set by an assumed equatorial
symmetry. The black hole singularity is handled by excising
a region inside the apparent horizon, i.e.\ by not placing colocation points
there.  The fluid grid does have points inside the excised region, but the fields
at these points are not evolved, and one-sided stencils are used to evolve points next
to the excision boundary.

The fluid is described by the baryonic rest mass density $\rho$, temperature $T$,
electron fraction (electrons per baryon) $Y_e$, and 3-velocity $v^i$.  The
pressure $P$ and specific internal energy $\epsilon$ are computed from
$\rho$, $T$, and $Y_e$ using an EOS table (see
Section~\ref{sec:EOS} below).  From these, one can compute the specific
enthalpy $h=1+\epsilon+P/\rho$.  From $v^i$, one can compute the Lorentz
factor $W=\alpha u^t$, $\alpha$ being the lapse and $u^t$ being the time
component of the 4-velocity.  The hydrodynamic evolution equations take
the form of conservation laws for a set of `conservative' variables.  These
include a density variable, $\rho_*=\sqrt{\gamma}W\rho$, an energy variable,
$\tilde\tau=\rho_*(hW-1)-\sqrt{\gamma}P$, a momentum variable
$\tilde S_i=\rho_*hu_i$, and a composition variable $\rho_*Y_e$, where
$\gamma$ is the determinant of the 3-metric, and $u_i$ are the covariant
components of the 4-velocity.  Note that we do not include neutrino
pressure or energy in the definitions of $\tilde\tau$ and $\tilde S_i$;
the neutrino radiation field appears only through source terms in the
evolution equations (see Section~\ref{sec:leakage}).

To carry out a timestep, we begin by computing the conservative variable
fluxes on cell faces using 5th-order WENO reconstruction \citep{Liu1994200,
Jiang1996202} of $\rho$, $T$, and $u_i$ and an HLL approximate Riemann solver
\citep{HLL}.  From these
cell boundary fluxes, the conservative variables are advanced forward in time. 
From the evolved conservative variables, the `primitive' variables $\rho$,
$T$, and $v^i$ must then be recovered.  This amounts to a two-dimensional
root-finding problem to find the values of $W$ and $T$ that reproduce the
evolved conservative variables.  For some combinations of conservative
variables, no corresponding set of primitive variables exists.  For this
reason, relativistic
hydrodynamics codes often impose the condition 
$\tilde S^2 < \tilde\tau (\tilde\tau + 2\rho_*)$~\citep{Etienne:2007jg}.
This inequality assumes $h\ge 1$ for all physical values of $\rho$ and $T$. 
One feature of nuclear-based EOSs is that the internal energy becomes
negative at sufficiently low densities and temperatures.  For an EOS
with a specific enthalpy minimum of $h_{\rm min}$, the invertibility condition
becomes
\begin{equation}
\tilde S^2 < \tilde S^2_{\rm max} = \tilde\tau (\tilde\tau + 2\rho_*)
+ (1-h^2_{\rm min})\ .
\end{equation}
To avoid divisions by zero, we impose a density floor of
$6 \times 10^3\rm\,g\,cm^{-3}$.
Gas near the surface of the matter, at points
with densities more than a few decades below the maximum, cannot
be expected to be evolved accurately; high temperatures and velocities tend
to develop there. We therefore impose a maximum value on
$T$ and $u^2 \equiv \gamma^{jk} u_j u_k$ in the low-density region outside the
star, tail, and disk (as in \citealt{Foucart:2013a}, but with modifications).
The ceiling on $T$ is the minimum temperature in the EOS table at densities below
$10^{-5}\rho^{\rm max}(t)$, where $\rho^{\rm max}(t)$ is the instantaneous maximum
density. We smoothly taper this treatment to densities 10 times larger than the lower
threshold by making the ceiling a linear function of density rising to 10~MeV.
Above $10^{-4}\rho^{\rm max}(t)$ no temperature modification is applied.
The ceiling on $u^2$ is 0 at densities below $10^{-7}\rho^{\rm max}(t)$,
and is smoothly tapered to 1000 at densities 100 times larger than the lower threshold.
Above $10^{-5}\rho^{\rm max}(t)$ no velocity modification is applied.
In addition, these density thresholds are increased by a factor of 100 in a region
very near the black hole. $Y_e$ evolution is not modified at any density.
By running segments of the evolution with different density
thresholds for the above treatments, we confirm that their main effect
is to reduce noise in the neutrino luminosity during tidal
disruption---noise that is dwarfed by the post-disruption signal anyway.

\subsection{Equation of State}
\label{sec:EOS}

In this work we describe the fluid with the Lattimer \& Swesty (hereafter LS) EOS.
This EOS is derived from a compressible
liquid-drop model with a Skyrme nuclear force and includes contributions
from free nucleons, alpha particles, and a single type of heavy nucleus \citep{Lattimer:1991nc}.
We set the nuclear incompressibility parameter, $K_0$, to 220~MeV and the
symmetry energy $S_v$ to 29.3~MeV.
Electrons, positrons, and photons are
added using the routines of \cite{1999ApJS..125..277T}.


We employ the LS EOS in our evolutions as a table of the following thermodynamic
and composition quantities: internal energy; pressure; sound speed;
neutron, proton, and electron chemical potentials;
and neutron, proton, alpha particle and characteristic heavy nucleus mass fractions
(along with the characteristic heavy nucleus's average mass and charge number).
We store the table with the following ranges and resolutions:
$\rho \in [10^{8},10^{16}]\rm\,g\,cm^{-3}$,
	$N=250$, log spacing;
$T \in [0.01,251]$~MeV,
	$N=120$, log spacing;
$Y_e \in [0.035,0.53]$,
	$N=100$, linear spacing.
As in \cite{OConnor2010} we perform tri-linear interpolations
for intermediate values. The original table and access routines
are available at \url{http://www.stellarcollapse.org}.
Continuous extrapolations of $P$ and $\epsilon$ outside the tabular bounds
are defined for primitive variable recovery. Note, however, that if $T$ or $Y_e$ at a
fluid grid point evolves to a value outside the table bounds, it is reset to the minimum or maximum.
Thus, becaue of this resetting, the extrapolations used for these
variables cannot influence the evolution.
However $\rho$ is allowed to take any value above the floor density. 
For densities above the floor density and below the table minimum, we
set the (negligible) pressure using a polytropic law with constants chosen
for each $T$ and $Y_e$ to smoothly match the low-density bound of the table. 
Densities greater than those
covered by the LS table do not occur in our simulation.

Just below nuclear saturation density a phase transition occurs
between nuclei at low densities and uniform nuclear matter at high densities.
The LS EOS captures this transition smoothly in free energy, but other parameters suffer from
jumps between the two phases. During the inversion from conservative to primitive variables,
we solve for temperature an equation involving enthalpy and pressure.
The inversion, therefore, is difficult
when either $h(T)$ or $P(T)$ are not smooth. We find that the sharp transition
is most disruptive to inversion at densities near $\rho \sim 10^{14}\rm\,g\,cm^{-3}$,
and temperatures near $T \sim 1$~MeV.
We resolve the issue by `polishing' the table in the temperature dimension,
applying a gaussian smoothing kernel of width 0.05~MeV to the stored values
of internal energy and pressure.

\subsection{Initial data}
\label{sec:ID}

We choose initial parameters that we expect to yield a massive accretion disk,
and high accretion efficiencies, giving us an upper bound on energetics.
The binary is characterized by a low mass ratio of $q=4$,
a high prograde aligned spin of $a^{*}=0.9$,
and an initial orbital separation yielding 6 orbits of inspiral before disruption.
In addition, we set the star's initial temperature to 0.01~MeV, and choose
a $Y_e$ profile that enforces $\beta$-equilibrium with zero neutrino
chemical potentials ($\mu_{\nu_i}=0$).
The star's baryonic rest mass is $1.55\,M_\odot$.  In isolation, it
would have a gravitational mass of $1.40\,M_\odot$ and an areal radius
of $12.7$~km.

To compute initial data we solve the extended conformal thin sandwich equations
using multi-domain pseudospectral methods, as detailed in
\cite{FoucartEtAl:2008}, using the Spectral Elliptic Solver
\citep{Pfeiffer2003a,Pfeiffer2003,Pfeiffer2004}.
The numerical domain is decomposed into a set of touching but not overlapping
`subdomains', comprised of spherical shells, filled spheres, cylinders,
and distorted cubes, arranged to reflect the symmetries in the configuration.
We use the same method to represent
the metric in evolutions (see e.g.\ Figures~1 and 2 in~\citealt{Foucart:2013a}).
One difficulty arising within a spectral framework
is the spurious Gibbs oscillations that can occur at discontinuities,
for example at the surface of the star.
Thus, a critical technique in our method involves capturing the
surface at a subdomain boundary, where the abutting spectral domains
have no difficulty representing the non-smooth field.
For the polytropes used in our previous
work~\citep{Duez:2009yy,Foucart:2010eq,FoucartEtAl:2011,Foucart2012,Foucart:2013a}
the density profile is smooth and
easy to resolve.  However new difficulties present themselves
with the complexity of the stellar structure derived from the LS EOS.

Polytropes have density profiles that
fall off as $\rho \propto (r-r_{*})^n$ close to the surface~\citep{Gundlach:2009ft},
where $r_{*}$ is the stellar radius, and
$n$ is related to the polytropic index by $\Gamma \equiv 1+1/n$.
Thus, density profiles of $\Gamma=2$ polytropes behave linearly at the surface.
To understand the effect of realistic EOSs on stellar surfaces
we may use the effective adiabatic index, defined
$\tilde{\Gamma} \equiv (d\log P/d\log \rho)_S$.
The LS EOS, at low temperature and $Y_e$ in $\mu_{\nu_i}=0$ $\beta$-equilibrium,
is extremely stiff near the central density of a $1.4\,M_{\odot}$ star, with
$\tilde{\Gamma} \sim 7/2$.
It begins to soften dramatically at $\sym10^{14}\,\text{g}\,\text{cm}^{-3}$,
to $\tilde{\Gamma} \sim 1/2$.
At even lower densities $\tilde{\Gamma}$
asymptotically approaches the adiabatic index of a relativistic
Fermi gas, $\tilde{\Gamma} \sim 4/3$.
These transitions occur very near the stellar surface,
at $r \gtrsim 0.95 r_{*}$ in our evolution coordinates.
The dramatic change in $\tilde{\Gamma}$ presents a kink in the density profile
that is difficult to resolve, especially with spectral methods.
Because the kink occurs over a range of radii,
it cannot be captured by a subdomain boundary,
as the well-defined surface of a polytrope can.

We can often improve the performance of our initial data solves
by manipulating the EOS minimally.
We find we can bypass most of the non-smoothness of the stellar profile
by appending a simple polytrope to the low-density portion of the cold EOS.
This makes the initial star more amenable to spectral representation.
Specifically, we use $P(\rho)=\kappa_0 \rho^{\Gamma_0}$
to describe pressures and
\begin{equation}
\epsilon(\rho) = \frac{\kappa_0}{\Gamma_0-1} \rho^{\Gamma_0-1} + \epsilon_{\rm shift} \nonumber
\end{equation}
to describe energies below $\rho_{\rm break}$.
$\rho_{\rm break}$ is determined by enforcing continuity in the effective
adiabatic index ($\tilde{\Gamma}(\rho_{\rm break})=\Gamma_0$),
$\kappa_0$ by enforcing continuity in pressure,
and $\epsilon_{\text{shift}}$ by enforcing continuity in energy.
We use $\Gamma_0=2$ with the LS EOS at $T=0.01$~MeV and $Y_e$ in
$\mu_{\nu_i}=0$ $\beta$-equilibrium.
This implies $\rho_{\rm break} = 4.49 \times 10^{13}\,\rm g\, cm^{-3}$,
$\kappa_0 = 1.40 \times 10^{4}\,\rm g^{-1}\,cm^{5}\,s^{-2}$,
and $\epsilon_{\rm shift} = 9.13 \times 10^{17}\,\rm erg\, g^{-1}$.

The effects of this manipulation on the bulk of the star are minimal.
It is helpful to bear in mind that the central density of a $1.4 M_{\odot}$
star described by this EOS is $\sym 7 \times 10^{14}\,\text{g}\,\text{cm}^{-3}$;
we append the polytrope more than an order of magnitude below this density.
Though the star's radius decreases by several percent, its rest mass changes
by one part in $10^5$. Test evolutions of isolated stars show that perturbations
from equilibrium, due to switching from the modified EOS used in the initial data
to the true EOS used in the evolution, are smaller than the numerical noise at
our typical resolutions.

In practice the manipulated density profile still falls off somewhat
non-smoothly at the surface, and we can sometimes gain further
accuracy in our initial data by adding an additional subdomain, isolating
the outer $20\%$ of the star (in radius) to a thin spherical shell.

Note that we limit these two surface-capturing methods to the initial data solve,
during which the metric \emph{and} fluid variables are represented on spectral subdomains.
During the evolution, the fluid variables are represented on a finite-difference
grid, where we also employ a high-resolution shock-capturing technique that
was designed to describe fluid discontinuities.

Because the quasi-equilibrium approximation ignores the infall motion
that should be present initially, using this data directly results
in eccentric orbits, which could affect the disk and outflow masses.
Therefore, we perform one iteration of eccentricity reduction (see
\citealt{Pfeiffer-Brown-etal:2007} for details) to incorporate the initial
infall and to reduce the initial eccentricity to $\sym0.01$.

\subsection{Neutrino leakage scheme}
\label{sec:leakage}
Neutrino leakage schemes attempt to account for
the energy loss and electron number alteration caused
by the emission of neutrinos by the nuclear fluid.  To accomplish this, one
first estimates the local effective energy emission rate
$Q_{\nu}$ ($\rm erg\,s^{-1}\,cm^{-3}$), the effective lepton number
emission rate $R_{\nu}$ ($\rm s^{-1}\,cm^{-3}$), and the neutrino
radiation pressure $P_{\nu}$ as measured in a local
Lorentz frame comoving with the fluid.  In such a frame
\begin{eqnarray}
\label{LLF1}
\partial_t (\rho Y_e) &=& -R_{\nu}m_U\ , \\
\partial_t T^{00} &=& -Q_{\nu}\ , \\
\label{LLF2}
\partial_t T_{j0} &=& -\partial_j P_{\nu}\ ,
\end{eqnarray}
where $m_U$ is the atomic mass unit,
and $T^{\alpha\beta}$ is the stress tensor.
Then the energy,
momentum, and composition source terms
follow directly from putting the above in generally
covariant form.  In the comoving Lorentz frame of
Equations~\ref{LLF1}-\ref{LLF2}, $u^0=1$, $u^j=0$,
and $g_{\alpha\beta}=\eta_{\alpha\beta}$,
so a covariant form of the neutrino sources is
\begin{eqnarray}
  \nabla_{\alpha}(\rho Y_e u^{\alpha}) &=& -R_{\nu}U\ , \\
  \nabla_{\alpha}T^{\alpha\beta} &=& -Q_{\nu}u^{\beta}
  - (g^{\alpha\beta}+u^{\alpha}u^{\beta})\nabla_{\alpha}P_{\nu}\ .
\end{eqnarray}
For the time derivative of $P_{\nu}$, we assume simple
advection $u^{\alpha}\nabla_{\alpha}P_{\nu}=0$, which is
not quite right, since the neutrino field changes in ways other
than advection, but the effect of the $\partial_tP_{\nu}$
source in our simulations (and, indeed, the effect of all neutrino
pressure terms) is quite small, so a more accurate
estimate is not necessary.

For the computation of $Q_{\nu}$ and $R_{\nu}$, our code is
essentially an extension of the GR1D neutrino leakage
code~\citep{OConnor2010} to three dimensions.  That code in turn closely
follows the leakage schemes of \cite{Ruffert1996} and
\cite{Rosswog:2003rv}.  In optically thin regions,
emission rates should be given by the local rates of
neutrino-generating interactions for each neutrino species
$\nu_i$:  $Q_{\nu_i}^{\rm free}$ and $R_{\nu_i}^{\rm free}$.  In
optically thick regions, energy and leptons escape via diffusion
at rates $Q_{\nu_i}^{\rm diff}$ and $R_{\nu_i}^{\rm diff}$.  The effective
rates come from interpolating these; e.g., for energy emission,
\begin{equation}
  Q_{\nu_i} =
  \frac{Q_{\nu_i}^{\rm free}Q_{\nu_i}^{\rm diff}}
       {Q_{\nu_i}^{\rm free}+ Q_{\nu_i}^{\rm diff}}\ .
\end{equation}
Like most other leakage schemes, we include three neutrino species,
$\nu_e$, $\overline{\nu}_e$, and $\nu_x$, where $\nu_x$ includes
$\mu$ and $\tau$ neutrinos and antineutrinos.  Then
$Q_{\nu}=Q_{\nu_e} + Q_{\overline{\nu}_e} + 4 Q_{\nu_x}$ and
$R_{\nu}=R_{\nu_e}-R_{\overline{\nu}_e}$.

For free emission rates, we include $\beta$-capture processes
($e^{-}p \rightarrow n\nu_{e}$ and $e^{+}n \rightarrow p\overline{\nu}_{e}$),
electron-positron pair annihilation, plasmon decay, and
nucleon-nucleon Bremsstrahlung emission.  (For the latter process,
we use the rate given in \citealt{Burrows2006b}.)  In calculating the opacity,
we include scattering on nucleons and heavy nuclei and absorption
on nucleons.  All particle species, including the neutrinos, are
assumed to have Fermi-Dirac distribution functions.

As in GR1D, we assume the neutrino chemical potentials
used to calculate the emission rates are
\begin{equation}
\label{mu_nu}
\mu_{\nu_i} = \mu_{\nu_i}^{\rm eq}(1-e^{-\langle\tau_{\nu_i}\rangle})\ ,
\end{equation}
where $\mu_{\nu_i}^{\rm eq}$ is the $\beta$-equilibrium value of
$\mu_{\nu_i}$ ($\mu^{\rm eq}_{\nu_e}=-\mu^{\rm eq}_{\overline{\nu}_e}
=\mu_e+\mu_p-\mu_n$, $\mu^{\rm eq}_{\nu_x}=0$) and $\langle\tau_{\nu_i}\rangle$ is
the energy-averaged optical depth.

Given the primitive variables
and neutrino chemical potentials, we can compute $\tau_{\nu_i}(E_{\nu_i})$, the
optical depth of neutrino species $\nu_i$ at energy $E_{\nu_i}$
along some path parameterized by $\ell$, with the line integral
\begin{equation}
\label{tau_integral}
\tau_{\nu_i}(E_{\nu_i}) = \int [\lambda_{\nu_i}(E_{\nu_i})]^{-1} d\ell
 = E_{\nu_i}^2 \int \hat{\lambda}_{\nu_i}{}^{-1} d\ell\ ,
\end{equation}
where $\lambda_{\nu_i}(E_{\nu_i})\equiv E_{\nu_i}{}^{-2}\hat{\lambda}_{\nu_i}{}^{-1}$
is the mean free path, and we have taken advantage of the fact that all
the absorption and scattering cross sections considered have a common neutrino
energy dependence, $\sigma_{\nu_i}\propto E_{\nu_i}^2$, which can be factored out.
Using the optical depth, the diffusion rates
$Q_{\nu_i}^{\rm diff}$ and $R_{\nu_i}^{\rm diff}$ are estimated as in
\cite{Rosswog:2003rv}.

The calculation of $\tau_{\nu_i}$ uses interaction cross sections
that depend on $\mu_{\nu_i}$, and by Equation~\ref{mu_nu}, these
themselves depend on $\langle\tau_{\nu_i}\rangle$.   GR1D iterates
Equation~\ref{mu_nu}, but since
optical depth calculations are more expensive in 3D, and the
interpolation formula is itself somewhat arbitrary, we have experimented
with introducing a second optical depth variable
$\langle\tau_{\nu_i}\rangle_{\rm approx}$ for use in Equation~\ref{mu_nu}. 
We have tried
\newline
1) setting $\langle\tau_{\nu_i}\rangle_{\rm approx}$ equal to
the energy average of $\tau_{\nu_i}(E_{\nu_i})$ and, as in GR1D, iterating
to rough convergence
\newline
2) limiting the code to one iteration per opacity calculation, starting
from previous calculation's results for $\langle\tau_{\nu_i}\rangle_{\rm approx}$
and $\mu_{\nu_i}$.  Since the matter distribution changes little between
opacity computations, this method is very similar to the first.
\newline
3) using an analytic fit for neutron stars: 
\begin{equation}
\label{tau_nu_rho}
\log_{10}\langle\tau_{\nu_i}\rangle_{\rm approx}=0.96\,(\log_{10}\rho_{\rm cgs}-11.7)\ ,
\end{equation}
4) modifying the above to capture the effect of temperature on the
average neutrino energy, and hence the cross-section:
\begin{equation}
\label{tau_nu_rho_and_T}
\log_{10}\langle\tau_{\nu_i}\rangle_{\rm approx}
=0.96\,(\log_{10}\rho_{\rm cgs}-11.7)\left(\frac{T}{0.1{\,\rm MeV}}\right)^2\ .
\end{equation}
The effect of the assumption for $\langle\tau_{\nu_i}\rangle_{\rm approx}$
on the luminosity is
fairly small for all epochs of our simulation (of order 10\% or less),
and the runs below use the simplest scheme (3).  The effect on
the lepton number emission rate can be tens of percents at times,
meaning the choice of neutrino chemical potential may have some effect
on the composition evolution of the disk, an issue we intend to explore
more deeply in future work.  As a first attempt to gauge the importance
of this choice, we have carried out numerical experiments evolving our
post-merger accretion disk at one resolution using different
$\langle\tau_{\nu_i}\rangle_{\rm approx}$ assumptions, namely schemes 2 and 3. 
Switching from scheme 3 to 2 does produce transient behavior, during which time
$R_{\nu}$ drops significantly, as points near the neutrinosphere adjust
to the different opacity profile.  After a short time, $R_{\nu}$
recovers, and the subsequent evolutions track each other decently. 
The notable aspect of the disk $Y_e$ evolution, the deleptonization at late
times as the disk cools, is found in both schemes and does not
seem to be an artifact of the chemical potential choice.
It should be emphasized that
$\langle\tau_{\nu_i}\rangle_{\rm approx}$ is only used for Equation~\ref{mu_nu}. 
The diffusion rates are determined using $\tau_{\nu_i}$ as computed in
Equation~\ref{tau_integral}.

\begin{figure}
\includegraphics[width=8.2cm]{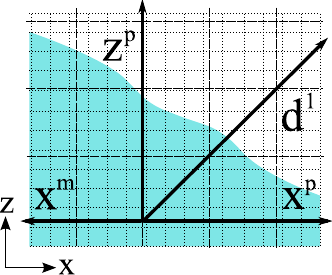}
\caption{
Schematic of the optical depth calculation. The fine (dotted) grid
represents the finite difference grid for the fluid evolution; the coarse
(dashed) grid represents the opacity grid for the optical depth calculation.
The blue region in the lower left represents dense fluid in which the neutrino
mean free path is small. For each zone of the opacity grid, we first integrate
five axial rays, excluding the $-z$
axis because of equatorial symmetry; three of these are labeled:
$\rm x^m$, $\rm x^p$, $\rm z^p$.
Then we integrate two additional `most promising' diagonal rays,
lying between the minimum optical depth axis and the two next-to-minimum axes;
one of these is labeled: $\rm d^1$.
}
\label{fig:chi_opacity_grid}
\end{figure}

The most costly part of the leakage scheme is the estimate of the
optical depth.  To compute this, we first interpolate the opacity
(factoring out the $E_{\nu}^2$ neutrino energy dependence, as discussed
above) onto a lower resolution 3D grid.  Then we compute
line integrals of this quantity in 7 directions to the computational boundary:
one integral forward and backward on each coordinate axis, excluding $-z$ because of our
grid's reflection symmetry, plus two diagonals based on the `most
promising' of the coordinate directions (see Fig.~\ref{fig:chi_opacity_grid}).
The minimum line integral gives the optical depth.
(More precisely, the optical depth is $E_{\nu}^2$ times this integral.)

To test our implementation of the leakage scheme, we have constructed
spherically symmetric configurations of hot gas and compared our
results ($R_{\nu_i}$, $Q_{\nu_i}$, and the optical depth) with those
of GR1D.  We set the neutrino chemical potentials in the same way for
this test to have a clean comparison.  We tried spheres with a range of
central density, temperature,
and $Y_e$, so that some were optically thin and others optically thick. 
As expected, the two codes agreed.

The emission rates also allow us to compute the total neutrino energy
and number luminosity radiated to infinity by performing proper integrals
over $Q_{\nu}$ and $R_{\nu}$.  For the energy luminosity at $r\rightarrow\infty$, $L_{\nu}$,
we multiply the integrand by a redshift factor $g_{00}$:  one factor of
$\sqrt{g_{00}}$ for the time dilation and one factor for the energy redshift. 
For the integral of $R_{\nu}$, we multiply by a redshift factor of
$\sqrt{g_{00}}$.

\section{Summary of evolution:  global measures and convergence}
\label{sec:grhydroconvergence}

\begin{figure*}[t] \centering
\includegraphics[width=18cm]{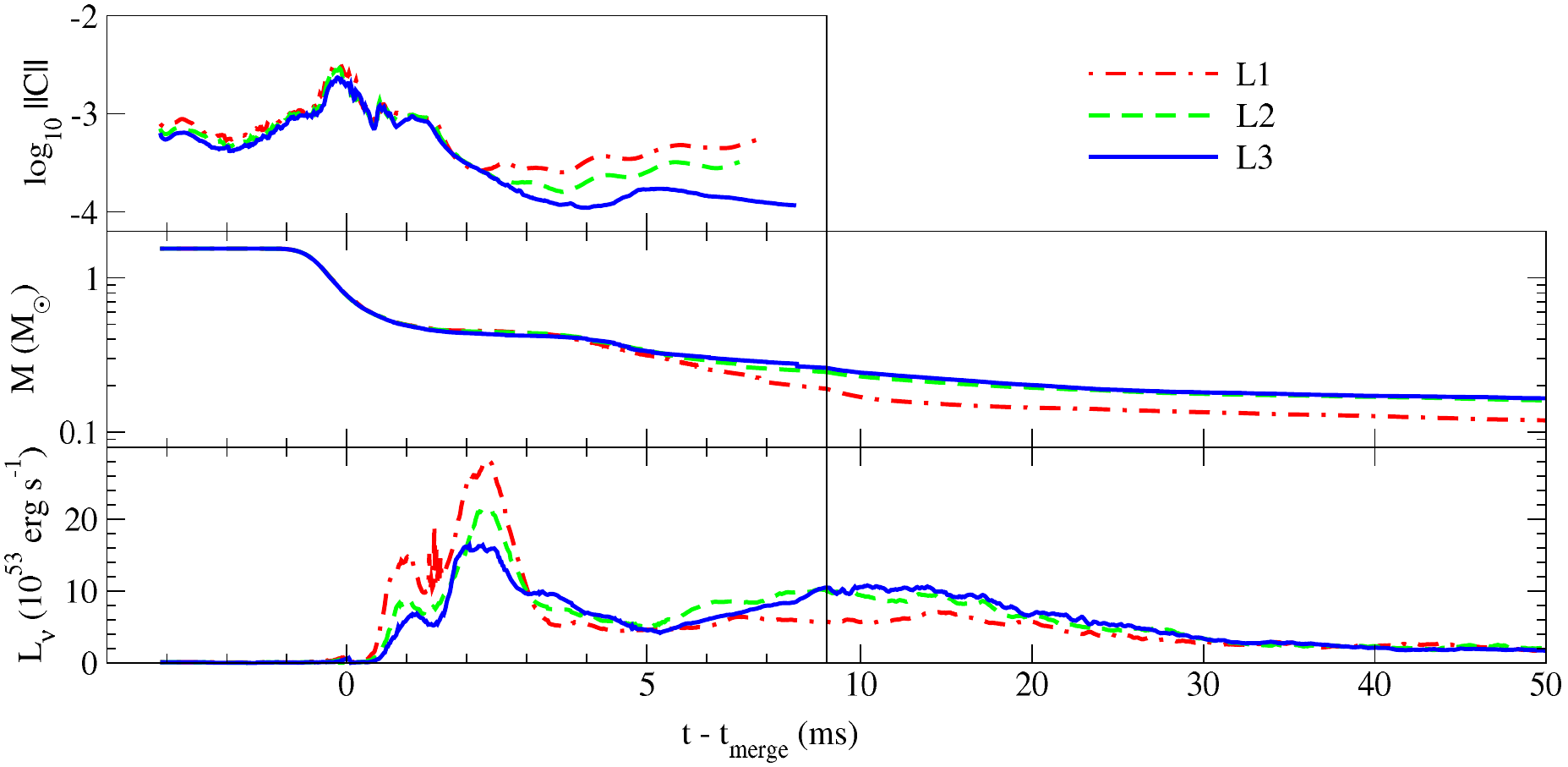}
\caption{
Global convergence measures,
showing the three resolutions with leakage.
$t_{\rm merge}$ is the time at which 50\% of the rest mass has depleted.
We use different scalings for the time-axis before and after 8~ms
in order to highlight the initial disruption and accretion.
{\em Top panel}: $L^2$ norm of global generalized harmonic constraint
violations (see \citealt{Lindblom:2007}). We stop tracking this measure at
$\sym7$~ms when we freeze the spacetime metric and begin the Cowling evolution.
{\em Middle panel}: baryonic rest mass remaining on the computational
grid (in units of $M_{\odot}$). The initial decrease beginning at $-1$~ms is
driven by accretion onto the black hole.
The second decrease beginning near $3$~ms is due to matter falling off the outer
bondary of the computational domain.
The disk mass quoted throughout ($M_0 \approx 0.3\,M_{\odot}$) is
the gravitationally bound mass outside the BH at $5$~ms in the L3 evolution,
i.e.\ the mass on the grid minus the instantaneous estimate of the remaining unbound mass
discussed in Section~\ref{sec:outflow}.
{\em Bottom panel}: total neutrino luminosity (in units of $10^{53}\rm\,erg\,s^{-1}$),
as measured at $r \rightarrow \infty$.
}
\label{fig:convergence}
\end{figure*}

Our evolution begins with a coordinate separation of 83~km
between the centers of the two compact objects.
The inspiral takes place over 6 orbits (28~ms) and is followed by tidal disruption and merger.
We continue to evolve the spacetime and fluid $\sym7$~ms past merger.
We define `merger' as the time when 50\% of the matter has
depleted, mainly by accretion (at this time, only a small fraction
($<10^{-3}$) has fallen off the outer boundary).
By 7~ms the metric is mostly stationary, but the accretion disk remains
highly dynamic.  The interpolation and communication of the fluid variables
to the spacetime (pseudospectral) grid is the dominant
computational cost during this phase of the evolution,
so continuing to evolve the complete system for many tens of
milliseconds would be prohibitive with the current code.  Instead,
we continue the evolution some 40~ms longer in
the Cowling approximation, that is we evolve the fluid but leave
the spacetime fixed.

The Cowling approximation ignores several effects.
First, it neglects changes in the disk's and the black hole's gravitational pull.
The mass accreted onto the black hole is some two orders of magnitude smaller than
the black hole mass, so this effect is probably unimportant.
Second, it neglects changes in the black hole's spin. We estimate the
change in the Kerr spin parameter through the 40~ms of Cowling evolution by
measuring the change in the angular momentum and mass of the fluid.
If we were to continue evolving the metric, $a^*$ would decay
(from $\sym0.9$) by 2\%, pushing out the radius of the innermost stable
circular orbit by less than 10\% \citep{1972ApJ...178..347B}.
We do not expect this effect to play a significant role in the disk evolution.
Third, the Cowling approximation cannot capture instabilities
due to the disk's self-gravity. However, this should not qualitatively
alter perturbations in the bulk of the disk.
The threshold for gravitational instability can be estimated from Toomre's
criterion:  $Q_T=\kappa c_s/(\pi G\Sigma) >1$ for stability
\citep{1960AnAp...23..979S,1964ApJ...139.1217T}, where $c_s$ is the sound
speed of the disk, $\kappa$ is the epicyclic frequency, 
$\Sigma$ is surface density, and $G$ is the gravitational constant.  For
the disk considered here, the minimum Toomre parameter there is about 20
except for the inner and outer edges.  The small amount of matter in
the outer regions of the grid is not yet in circular orbit equilibrium,
so the stability condition is inapplicable.  At the innermost region
of the disk, $\kappa\rightarrow 0$ at the innermost stable circular
orbit, but here the Cowling approximation does capture
the main orbital instability.
Fourth, the Cowling approximation discards the gravitational
waves caused by the disk's nonaxisymmetry.  For the observed mass
quadrupole variations induced in the disk, gravitational waves will
affect the modes on a timescale $E_{\rm mode}/L_{\rm GW}\sim 10^2$~s
(where $E_{\rm mode}\sim M_0 v^2(\delta\Sigma/\Sigma)^2$ is a
characteristic energy of the spiral waves, and
$L_{\rm GW}$ is the gravitational wave luminosity).
Thus we may safely ignore this effect.
Finally, this approximation `freezes in' any nonaxisymmetric modes present
in the metric when we transition to the Cowling evolution.
We see that the amplitude (normalized to the $m=0$ mode) of the $m\geq1$ modes
in the lapse are below $\sym10^{-3}$ at all radii. We expect this
nonaxisymmetry to feed back into the fluid at the same order of magnitude.

To check the accuracy and robustness of our results, we evolve
the merger at 3 resolutions, which we label 'L1',
'L2', and 'L3'.  See Table~\ref{tab:resolutions} for a
summary of the resolutions.  Since the
spectral evolution uses adaptive meshing, the number of colocation
points changes significantly over the course of an evolution, so
we report average resolutions.  The actual numbers of stored fluid
grid points are half those shown in the table, since each stored number
gives the fluid variables at a point below and a point above the
equator, following our assumption of equatorial reflection symmetry. 
After tidal disruption, the coordinate system of our fluid grid
is driven nonuniformly to higher resolution close to the black hole.
Our high resolution corresponds to a fluid grid spacing of $\sim$160~m during
inspiral and about 0.7~km vertical / 2~km radial in the inner
disk after merger.

\begin{table}
  \centering
  \begin{tabular}{ l l l l l }
    \toprule
                    &                             & L1        & L2        & L3 \\
    \midrule
    $N$ gridpoints  & spectral domain             & $70^3$    & $75^3$    & $87^3$ \\
                    & hydro domain                & $120^3$   & $140^3$   & $160^3$ \\
    \midrule
    $\langle dx \rangle$ (km) & early: star       & 0.21      & 0.18      & 0.16 \\
                    & late: inner disk (vert/rad) & 0.9 / 2.7 & 0.8 / 2.3 & 0.7 / 2 \\
    \bottomrule
  \end{tabular}
  \caption{
    Summary of numerical resolutions. $N$ is the total number of gridpoints in the
    evolution domains. $\langle dx \rangle$ is the average grid spacing on the hydro
    domain, calculated in evolution coordinates.
  }
  \label{tab:resolutions}
\end{table}

In addition to the convergence tests, we perform one merger, at the
L1 resolution, without neutrino leakage.  We label this
run `L1no$\nu$'.  Comparing L1no$\nu$ to L1 lets us isolate
the effects of the neutrinos on the fluid.

\begin{figure*} \centering
\includegraphics[width=18cm]{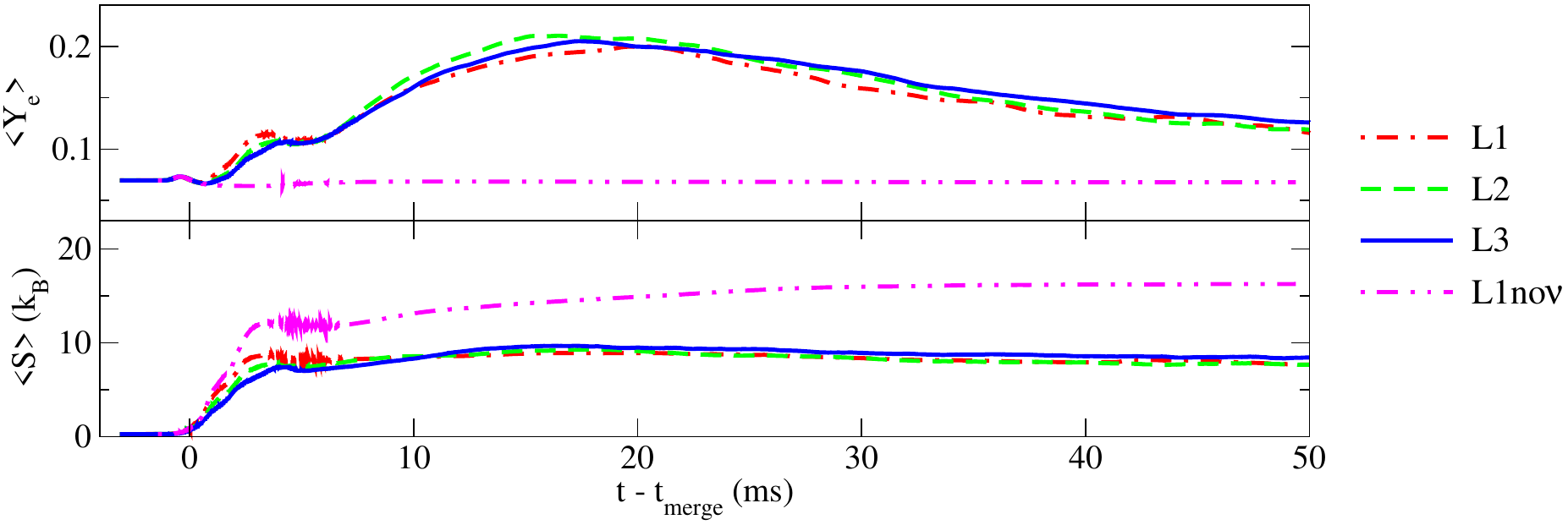}
\caption{
Secular chemical and thermal evolution of the disk for the three resolutions
with leakage, and also L1no$\nu$.
{\em Top panel}: mass-weighted average of the electron fraction.
In the L1no$\nu$ evolution, $Y_e$ simply advects with the flow.
The early decreases at 0~ms and 3~ms are due to the major mass loss episodes:
initial accretion, and loss of the tidal tail from the outer boundary.
{\em Bottom panel}: mass-weighted average of the entropy, in units of
$k_B\rm\,baryon^{-1}$.
}
\label{fig:globalevolution}
\end{figure*}

In Figures~\ref{fig:convergence} and~\ref{fig:globalevolution},
we show the evolution of several main global quantities for all runs. 
Comparing resolutions allows us to discern
which evolution features are robust and which are numerical artifacts. 
For example, the neutrino luminosity ($L_{\nu}$) spikes 2~ms after merger,
but this feature appears to converge away as resolution is increased.
Volume renderings of the fluid (see an equatorial slice in Fig.~\ref{fig:disruptionsnapshots})
show that at these times, some of the nuclear matter is spread in
a thin tidal tail which is underresolved at low resolutions. The luminosity
at later times comes from the much better-resolved disk, and we see
good agreement between the higher resolutions on $L_{\nu}$ after
about 5~ms.
Figure~\ref{fig:convergence} reveals that the rest mass
of L1 at late times differs from that of the higher resolutions significantly
more than the high resolutions differ between themselves.
By 10~ms this difference has settled to about 30\%.
The reason for this is that resolution L1 generates a hotter, fatter,
lower-density disk.
The puffed-up disk drives more of its matter off of our computational domain,
an effect that we can already see in outflow measurements by 2~ms after merger
(see Section~\ref{sec:outflow}).
Finally, in Fig.~\ref{fig:globalevolution}, we again see that the different
resolutions agree well on the post-merger composition and entropy,
giving us confidence that these are roughly correct, though
strict convergence is lacking.  Given the strong shocks and
turbulent-like behavior in the disk, this is unsurprising.
However, comparing the leakage runs to L1no$\nu$, we see that the secular
neutrino effects are much larger than the differences between resolutions.

\section{Tidal disruption and ejecta}
\label{sec:outflow}

In a BHNS merger, matter expelled at high velocity
may ultimately become unbound from the central gravitational potential.
In addition to enriching the interstellar medium with r-process elements
\citep{1974ApJ...192L.145L,Freiburghaus:1999no,Arnould:2007gh,korobkin:12},
this nuclear matter may emit an electromagnetic signal.
The radioactive nuclei formed in the neutron-rich fluid
quickly fission, emitting high-energy beta- and gamma-radiation.
This heats the ejecta, which becomes optically thin on an expansion
timescale (days to weeks), and radiates thermalized photons, producing
an isotropic kilonova (or `macronova', see
\citealt{Li:1998bw,Roberts2011,metzger:11,Rosswog:2012fn,
2013MNRAS.tmp..771P,Kasen:2013xka,NissankeEtAl:2012})
at optical, or perhaps infrared, wavelengths, depending on the highly-uncertain
opacity properties of the material \citep{Kasen:2013xka}.

Additionally, the high-velocity ejecta form a blast wave in the circumbinary medium.
At the shock front, magnetic fields amplify and accelerate electrons and positrons,
thus emitting synchrotron radiation.
This radio emission continues for a deceleration timescale (months to years)
dependent upon the density of the surrounding medium
\citep{metzger:11,Nakar:2011cw,2013MNRAS.tmp..771P,NissankeEtAl:2012}.

These transient electromagnetic signals are of sufficient interest
to warrant a thorough examination of any ejecta in our simulation.
There are two dominant conveyers of ejecta in BHNS mergers:
tidal tails and accretion disk winds.  In this simulation, most of
the unbound mass is produced by the tidal tail.  A smaller amount is
ejected in a plume during the early merger phase, as the infalling gas stream
collides and shocks with itself. Also a small amount of outflow from the late disk
is observed, but at low densities where numerical errors can introduce
spurious acceleration and heating of the fluid.
Note that this simulation ignores some effects that could drive
strong winds (e.g. neutrino heating) and does not evolve long enough to
see some others (e.g. large-scale He recombination).
Below we review previous results, describe the formation of the tidal tail,
describe our method of analyzing the ejecta, and finally report measures of
its mass ($M_{\rm ej}$), kinetic energy ($E_{\rm ej}$),
distribution of velocities, and composition ($\langle Y_e \rangle_{\rm ej}$).

There are a number of recent studies examining ejecta from
binary neutron star mergers (e.g.\ \citealt{Rosswog:2012fn,hotokezaka:13}),
high-eccentricity mergers (e.g.\ \citealt{Stephens:2011as,2012PhRvD..85l4009E,East:2012ww}),
and mergers including magnetic fields (e.g.\ \citealt{Chawla:2010sw}).
But here we survey results from low-eccentricity, nonmagnetized, BHNS studies
to emphasize the influence of black hole spin.
\cite{1974ApJ...192L.145L} made semianalytic estimates in a Schwarzschild spacetime.
They found $M_{\rm ej} \sim 0{\rm\,\,to\,\,}0.14\,M_{\odot}$ for stars that disrupt
close to the black hole; no mass is ejected from stars that disrupt far away.
\cite{Rosswog:2012fn} simulated two cases of $q\sim4$ and $q\sim7$ in a
Newtonian potential. He found $M_{\rm ej} \lesssim 0.05\,M_{\odot}$.
Relativistic simulations have characteristically yielded more
conservative ejecta estimates.
Notably, high mass-ratio, compact neutron star, low black hole spin systems do not even
disrupt. (See \citealt{Pannarale:2010vs} and \citealt{Foucart2012} for phenomenological models
covering this parameter space.)
In an excellent study focused on unique signatures from BHNS merger ejecta,
\cite{Kyutoku:2013wxa} simulated a suite of tens of mergers
with mass ratios of $q=3$ to $7$ and prograde BH spins up to $a^{*}=+0.75$.
They showed $M_{\rm ej} \sim 0.01\,M_{\odot}$ to $0.07\,M_{\odot}$, with more matter ejected
if the EOS is stiff.
Recently, \cite{Foucart:2013a} simulated several
mergers of $q=7$, with large BH spins of $|a^{*}|=0.9$ varying in inclination.
In cases of aligned spin and orbital angular momentum, they found ejected
masses of $M_{\rm ej} \sim 0.09\,M_{\odot}$.
Finally, in a study with nearly-extremal BH spin, \cite{Lovelace:2013vma} found
$M_{\rm ej}$ could be as high as $0.3\rm\,M_{\odot}$.
It appears, then, that for BHNS systems, one parameter region of interest for significant ejecta
may be the parameter region of high spin.

\begin{figure*} \centering
\includegraphics[width=18cm]{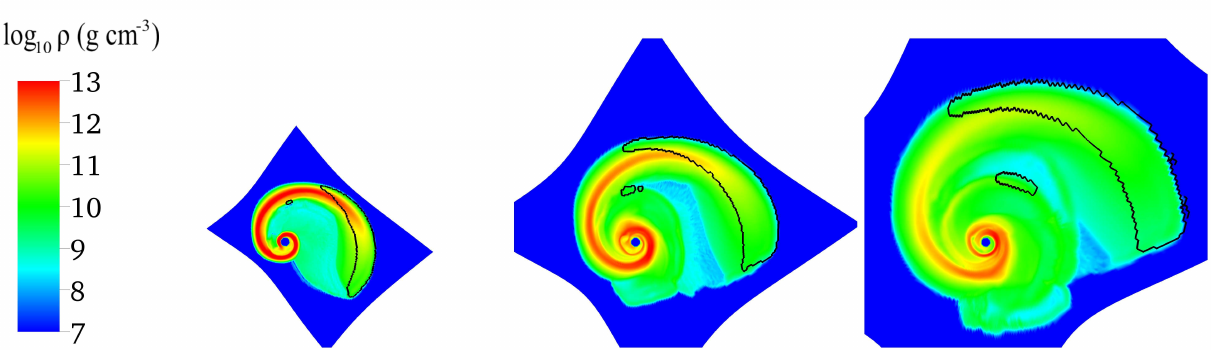}
\caption{
Equatorial slices of rest density from L3
during disruption and tidal tail formation at 1, 2, and 3~ms after merger
(from left to right).
The unbound material is outlined in black. The field of
view is fixed in all panels, so that each
has a scale of approximately $680\times680\rm\,km^2$.
The computational grid expands with the tail until it covers a circle of
radius $\sym400$~km (in our asymptotically flat evolution coordinates)
at which we remove matter from the grid. Note that we distort the
coordinate map between the fundamental evolution frame and the
frame in which the finite difference grid is uniform
in order to concentrate resolution near the black hole (see \citealt{Duez:2009yy}).
}
\label{fig:disruptionsnapshots}
\end{figure*}

In the present simulation, we see a large tidal tail form just before merger.
About two orbits before merger ($t=-3$~ms) the coordinate separation
of the two centers of mass has decayed to
$40$~km and the neutron star has become extremely distorted.
After another revolution ($t=-1.5$~ms) the separation has decayed to $25$~km,
the star overflows its Roche lobe, and it begins to accrete onto the black hole.
A tidal tail has already formed from the trailing edge of the star,
extending outward and lagging an orbit behind the core.
At $t=0$, the core falls into the black hole.
Over the next 3~ms the tidal tail sweeps out and
away from the black hole (see Fig.~\ref{fig:disruptionsnapshots}).
We follow the tail by periodically resizing our computational domain outward to
a radius of $\sym400$~km, where we allow the tail to fall off of the grid.
From its formation until its exit, the tidal tail is evolved on the computational domain
for $\sym5$~ms.

For a stationary spacetime, $u_t$ (the projection of the 4-velocity along
the timelike Killing vector field) is a constant of motion for geodesics. 
Assuming the space is also asymptotically flat, $u_t$ gives the Lorentz
factor ($\gamma=-u_t$) of the particles as they escape to infinity.  To the
approximation that the spacetime is settled and that pressure is dynamically
negligible, $u_t$ will be a constant along fluid parcels, and matter with
$u_t<-1$ may be flagged as unbound. Neither approximation is strictly true,
but both become better satisfied as the outflow expands outward away from
the dynamical region and decompresses to low pressures.  Thus, $u_t$ of
fluid elements in the tail, and the total amount of unbound material in
the tail, should become constant as the tail expands.  If this settling
happens before the tail leaves the computational grid, meaningful statements
about the amount of outflow and its asymptotic velocity distribution can
be made.

We integrate the mass satisfying the unboundedness
condition $u_t<-1$ over times from $t \sim 0$~ms, just as
the tidal tail is forming, to $t \sim 7$~ms, after the tail has entirely fallen off
the computational domain. We mask out regions within
$50\,M_{\odot}$ (73~km) of the BH in the evolution coordinates.
At this distance the Newtonian gravitational
potential is $M_{\rm BH}/r \lesssim 0.15$.
The integral of total mass ejected is robust against
30\% variations in the mask radius.
Furthermore, our velocity cap (see Sec. \ref{sec:Methods}) is only applied
at densities below $\sym 10^{8}\rm\,g\,cm^{-3}$ during this epoch, a full 3
orders of magnitude below the maximum density in the tail.

We check the validity of our assumption that the spacetime has become
stationary by estimating the timescale that $u_t$ will change due to changes
in the metric.  For a nonrelativistic geodesic, the leading order term for
the time derivative is $\dot{u}_t\approx g_{tt,t}$, so the timescale on which
metric nonstationarity can produce a large relative change in energy is
$\approx (-u_t-1)/\dot{u}_t$.  This is $\sim$~50~ms to 1~s for our ejecta
when near the edge of the computational domain.  On the other hand, the
metric is rapidly settling as the gravitational wavetrain overtakes the
ejecta, and $\dot{u}_t$ drops by an order of magnitude in a few ms.  Thus,
the ejecta energy is not expected to change by a large amount.  However,
tracking the ejecta for an extra few milliseconds would significantly
reduce this source of error, an important consideration for future
simulations.

\begin{figure}
\includegraphics[width=8.2cm]{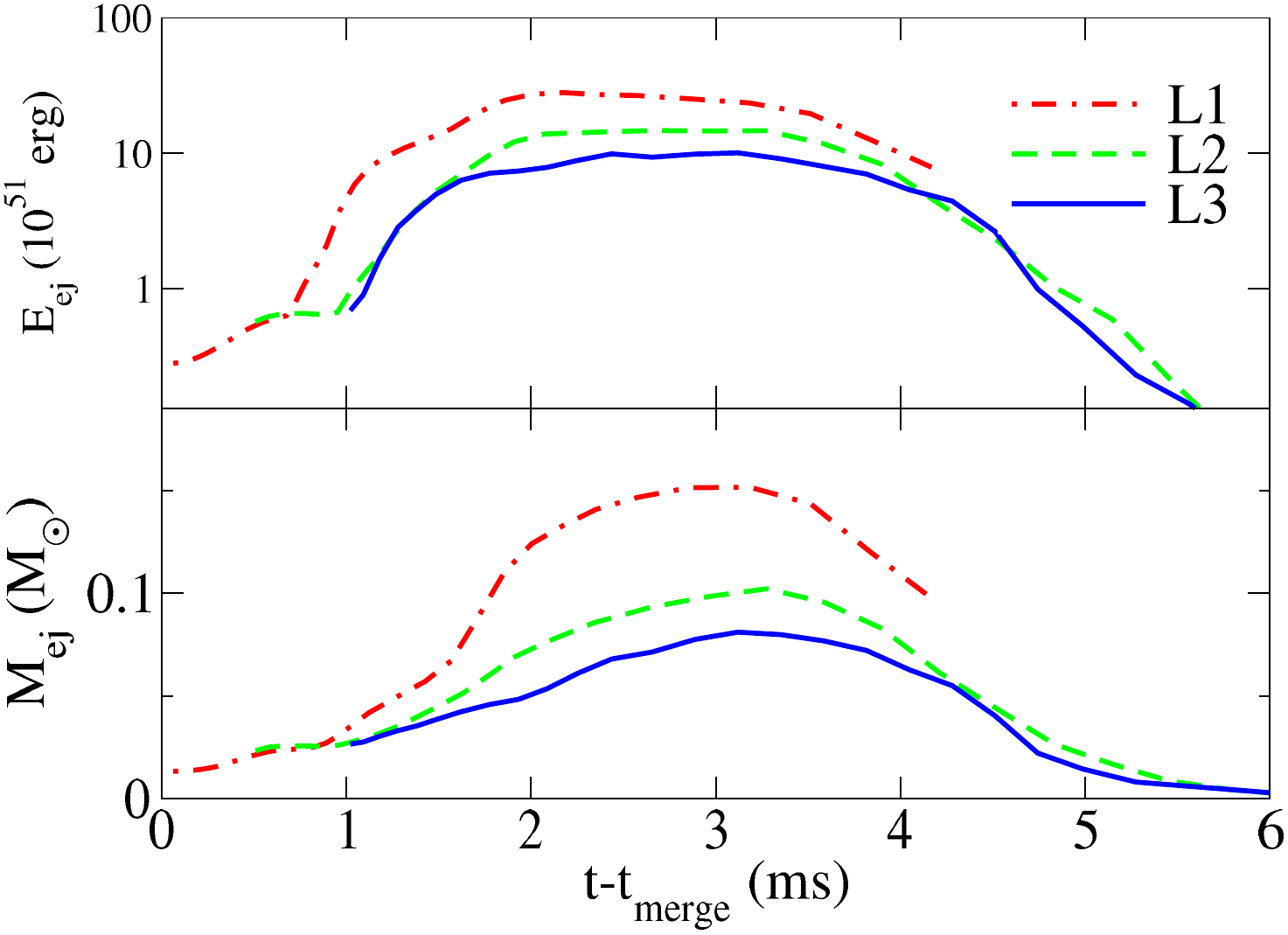}
\caption{
Integrals of the unbound fluid on the grid.
{\em Top panel}: total asymptotic kinetic energy.
{\em Bottom panel}: total rest mass.
For both integrals a sphere of 73~km near the black hole was masked out.
The drop-off after 3~ms is due to the loss of matter through
the outer boundary.
}
\label{fig:ejecta_mass_and_energy}
\end{figure}

Measures of $M_{\text{ej}}$ and $E_{\text{ej}}$ from the three resolutions
are shown in Fig.~\ref{fig:ejecta_mass_and_energy}, where we have calculated the total integral
over unbound matter on the grid.
This raw measure increases early as a growing amount of material
is flung out from the disk and exits the masked region.
In addition to the tail, volume renderings of the unbound matter reveal the existence of
a plume of outgoing matter ejected above the equatorial plane, of lower density than the tidal tail visible
in Fig.~\ref{fig:disruptionsnapshots}.  This second outflow is produced during
the merger, perhaps by the shock waves in the infalling matter. 
The most energetic matter in this plume begins to fall off the
computational domain after 2.5~ms, whereas the dense tidal tail begins to fall off
after 3~ms. Therefore, we use the peak values of mass and kinetic energy in
Fig.~\ref{fig:ejecta_mass_and_energy} as conservative estimates.
We use Richardson extrapolation to derive uncertainties from the highest two resolutions,
and we assume $2^{\rm nd}$ order convergence, though Fig.~\ref{fig:ejecta_mass_and_energy}
indicates approximately $6^{\rm th}$ order covergence. This method gives
$M_{\rm ej}= 0.08 \pm 0.07 \,M_{\odot}$, and
$E_{\rm ej} = 10 \pm 17 \times 10^{51}$~erg.
These errors are upper bounds which are probably large overestimates.
Nonetheless, even at our highest resolution, the tail is poorly resolved:
at $\sym70$~km wide, it is covered by $\sym15$ grid cells. This is a common limit
of grid-based methods; smoothed-particle hydrodyanmics methods neatly overcome this
limit, but find their handicap in evolving accurate initial conditions for the tail.

\begin{figure}
\includegraphics[width=8.2cm]{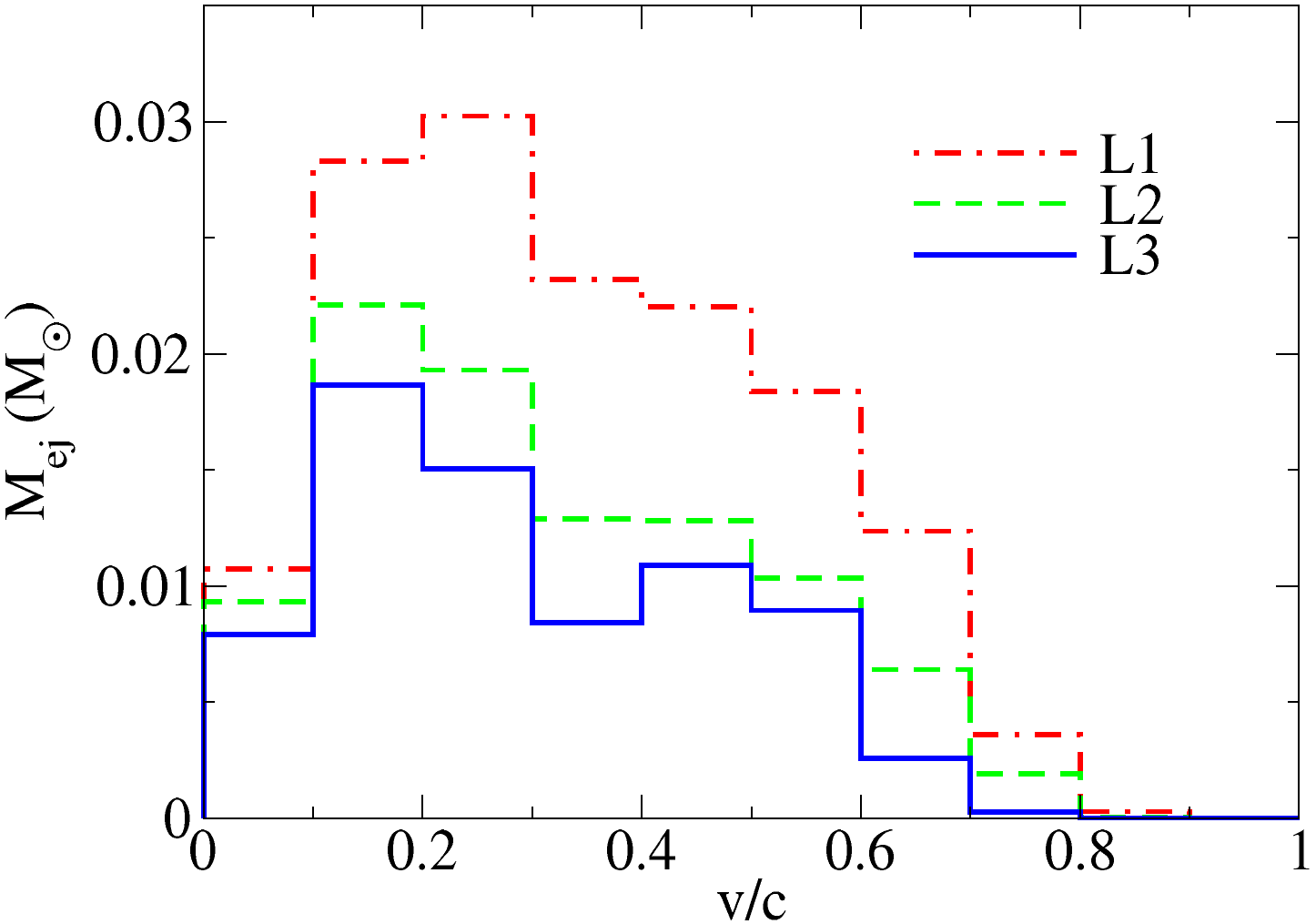}
\caption{
Unbound rest mass binned by magnitude of the asymptotic fluid velocity.
We show L1, L2, and L3 at $~\sym2.7$~ms, just before the peak in
Fig.~\ref{fig:ejecta_mass_and_energy}.
}
\label{fig:ejecta_velocity_histogram}
\end{figure}

The velocity distribution peaks across all resolutions at $0.2c$, but the dispersion in
velocities is large, with some matter escaping at $v>0.5c$
(see Fig.~\ref{fig:ejecta_velocity_histogram}).
Because average temperatures in the tidal tail are $\sym 1$~MeV as it leaves
the computational domain, the fluid is still in nuclear statistical equilibrium,
so our equation of state is still valid.
The material ejected from the
system is neutron-rich, with average electron fraction
$\langle Y_e \rangle_{\rm ej} \approx 0.1$.

\section{Relevant timescales for the disk}
\label{sec:scales}

The density in the accretion disk peaks at a circumferential
radius $r$ of around 50\,km from the black hole, and most of the
disk mass is within 100\,km of the black hole.  At the density
peak, the density is $\rho\approx 10^{12}$g cm${}^{-3}$, the
angular frequency $\Omega$ is around 3\,kHz, and the sound speed $c_s$
is around 0.1$c$.  The disk is hot and
thick:  $H/r\sim c_s/(\Omega r)\sim\frac{1}{3}$. 
The dynamical timescale of the disk is, to order of magnitude,
\begin{equation}
T_{\rm dyn} \sim \Omega^{-1} \sim \frac{H}{c_s} \sim \frac{r}{c_s}\
\sim 1\,\mathrm{ms}.
\end{equation}
Angular frequency decreases with distance from the black hole, so the
orbital period at $r=$120\,km (roughly speaking, the disk's outer edge)
is 10\,ms.  Thus, even this outer region is evolved for several
dynamical times. 
The perturbations in the inner disk rotate at frequencies similar to
the local flow and thus orbit the black hole on millisecond periods. 

The disk has strong entropy gradients in the inner and outer regions,
so disk gas can experience buoyancy forces there on a timescale given by the
Brunt-V\"ais\"al\"a frequency $N_{\rm BV}$,
\begin{equation}
T_{\rm S} \sim \left|N_{\rm BV}{}^2\right|^{-1/2}
=  \left|-\rho^{-1}(\partial\rho/\partial S)_P\,g\cdot\nabla S\right|^{-1/2} \approx 1{\rm\,ms}\ ,
\end{equation}
where $g=\rho^{-1}\nabla P$ and $S$ is the specific entropy. 

In addition to these dynamical processes, there are secular processes
that alter the disk structure on longer timescales.  One such effect
is the depletion of baryonic mass in the disk. 
The disk's mass is $M_0 \approx 0.3M_{\odot}$ after merger
(see Fig.~\ref{fig:convergence}).
For the first 30\,ms after merger, accretion onto the black hole proceeds at a rapid
rate of $\dot{M}_0 \approx 2\,M_{\odot}\rm\,s^{-1}$.  During this time, the
gas in the outer disk settles into circular orbit, which requires a
transfer of angular momentum from the inner disk.  (See
Section~\ref{sec:disk} for a fuller discussion of the disk's angular
momentum evolution.)  The transport of angular momentum away from the inner
disk by spiral density waves drives accretion onto the black hole. 
Eventually, the high-density middle region of the disk settles to
equilibrium, but dynamical, nonaxisymmetric distortions persist
near the disk's inner and outer edges (also discussed in
Section~\ref{sec:disk}).  Spiral waves thus travel outward and drive
a reduced rate of accretion throughout the simulation.  After 30\,ms, the
accretion rate has dropped by nearly an order of magnitude.  At
these late times, disk mass is also lost to a weak outflow from the inner
disk at a comparable rate to the accretion into the black hole.  Combining
these effects, we can define a mass depletion timescale
\begin{equation}
T_{\rm dep}\sim \frac{M_0}{\dot{M}_0}\gtrsim 0.2{\rm\,s}\ .
\end{equation}

The disk's thermal evolution is driven by shock heating, compression,
advection of heat into the black hole, and radiative cooling. 
The disk's thermal energy is defined as
\begin{equation}
E_{\rm thermal}=\int \rho_*[\epsilon - \epsilon_{\rm cold}(\rho,Y_e)]d^3x,
\end{equation}
where $\epsilon$ and $\epsilon_{\rm cold}(\rho,Y_e)$ are the actual
specific internal energy and the specific internal energy at the
lowest temperature in the EOS table, for which the gas is degenerate. 
We find $E_{\rm thermal}\approx 10^{52}\rm\,erg$. 
The total neutrino luminosity is $L_{\nu} \sim 10^{53\text{--}54}\rm\,erg\,s^{-1}$. 
Therefore, the cooling timescale due to neutrinos is
\begin{equation}
T_{\rm cool}\sim \frac{E_{\rm thermal}}{L_{\nu}} \approx 10{\rm\,ms}\text{--}100{\rm\,ms}.
\end{equation}
It is important to remember that the disk is partly pressure-supported.
Thus radiative energy loss
may come from the potential energy reservoir via disk contraction,
in addition to the more obvious thermal energy reservoir via disk cooling.
Indeed we find that evolving without neutrino cooling
leads to a disk that is not only hotter but also much more extended and
less dense.  The timescale for composition change
is $N_e/R_{\nu}$, where $N_e$ is the number of electrons in the disk and
$R_{\nu}$ is the total net lepton number change rate due to neutrinos. 
This is initially also about
10~ms.  Then, 20~ms after merger, balance between $\nu_e$ and $\overline{\nu}_e$
emission is roughly achieved, and the
composition subsequently changes more slowly.   Thus, the neutrino emission
significantly influences the energy and composition of the disk over its
lifetime.

\section{Accretion disk:  dynamical equilibrium and stability}
\label{sec:disk}

\subsection{Disk formation}

\begin{figure}
\includegraphics[width=8.2cm]{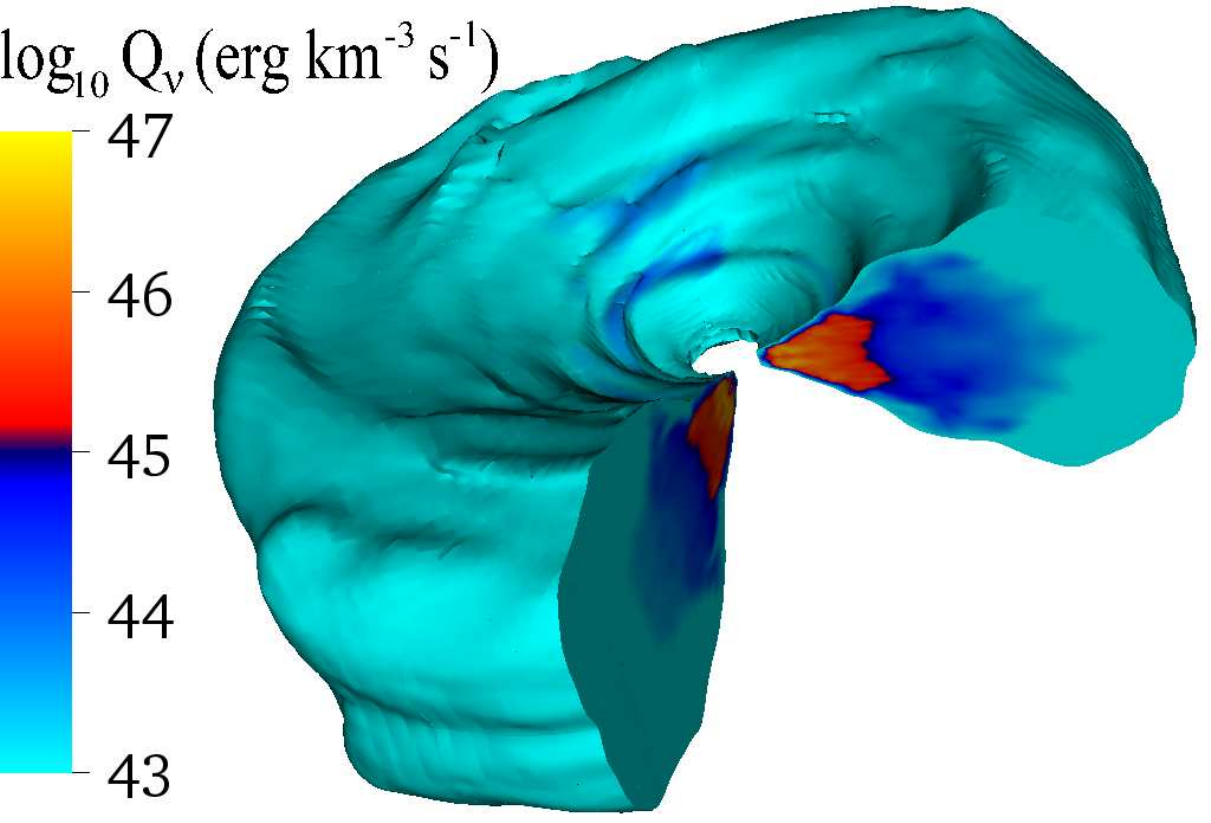}
\includegraphics[width=8.2cm]{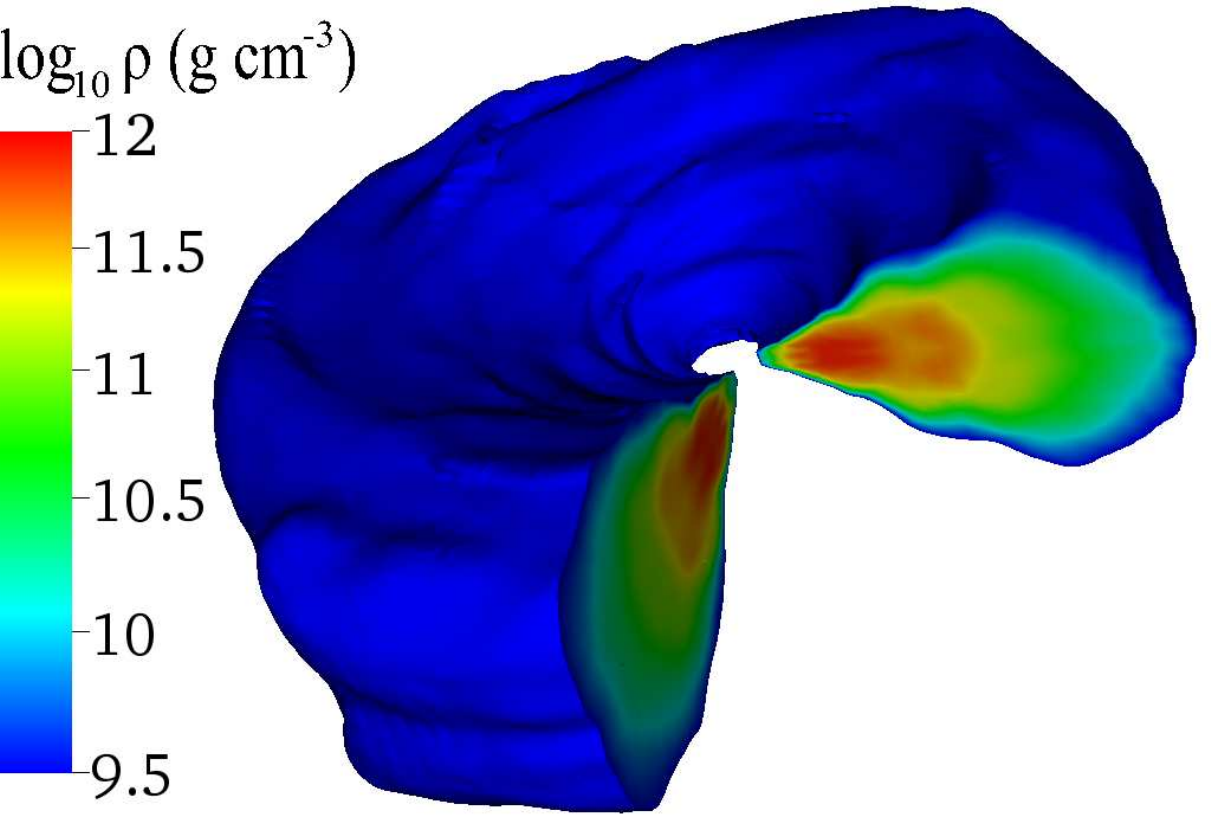}
\includegraphics[width=8.2cm]{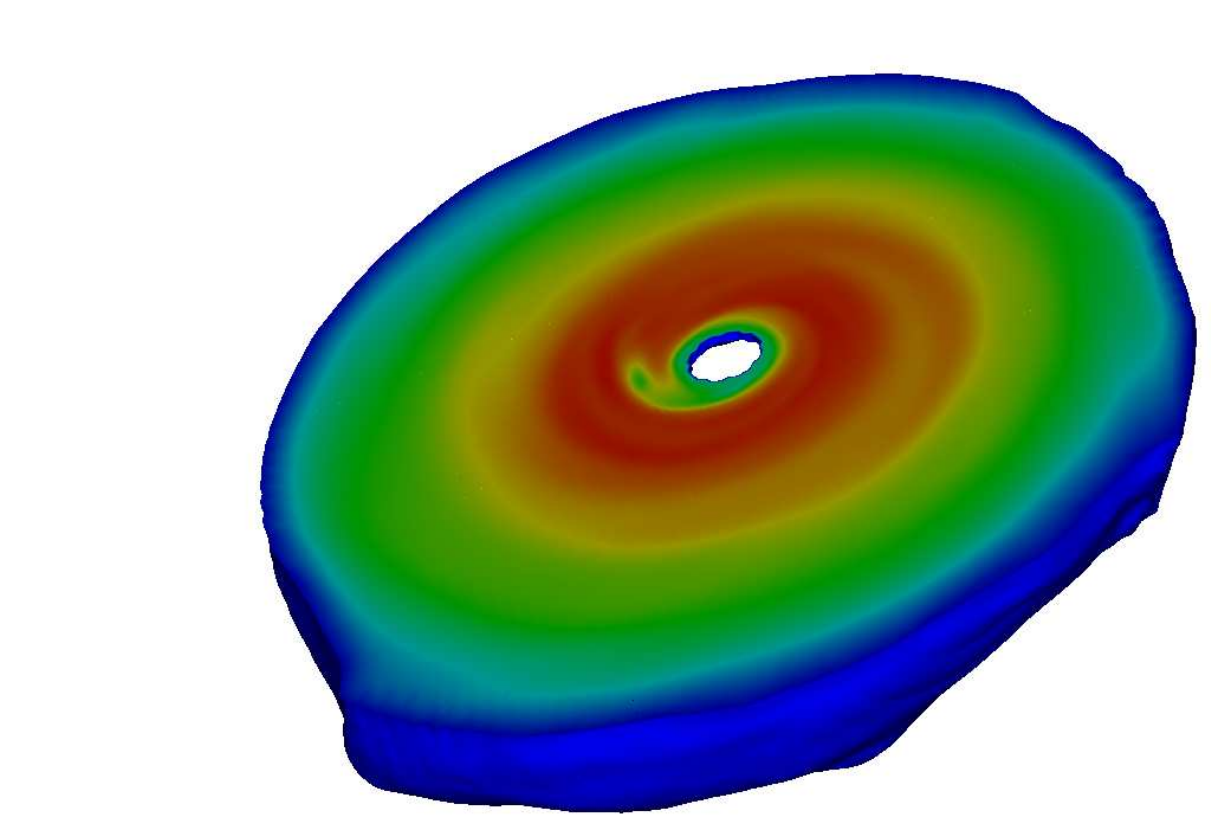}
\caption{
Three-dimensional distribution of neutrino energy loss and fluid density
from L3, 30~ms after merger, depicted in evolution coordinates at a slanted view.
{\em Top panel}: effective local neutrino power, $Q_{\nu}$ (no redshift applied),
summed over all species (in units of $\rm erg\,km^{-3}\,s^{-1}$).
{\em Middle and bottom panels}: meridional and equatorial slices of
density in the fluid rest frame. The equatorial slice reveals a spiral mode.
In all three panels, densities below $10^{9.5}\rm\,g\,cm^{-3}$ have been masked out
to show the structure of the disk. The disk radius and half-thickness
are 110~km and 45~km, respectively.
}
\label{fig:disksnapshot}
\end{figure}

\begin{figure}
\includegraphics[width=8.2cm]{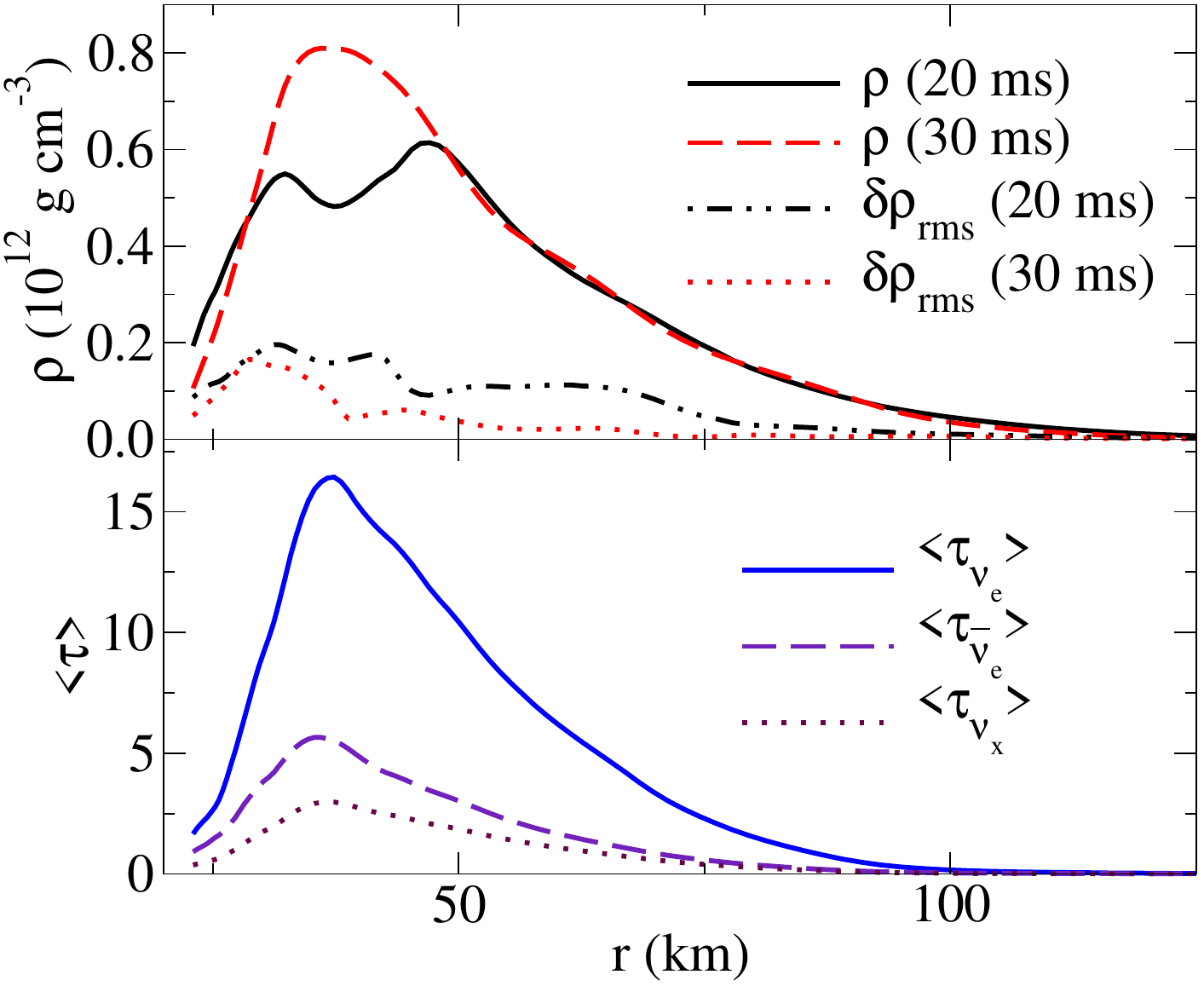}
\caption{{\em Top panel}: azimuthally-averaged profiles of the
density $\rho$ together with its RMS deviation from axisymmetry,
$\delta\rho_{\rm rms}\equiv \langle(\rho-\langle\rho\rangle)^2\rangle^{1/2}$,
plotted as functions of the circumferential radius
at the equator $r$.  Values for 20~ms and 40~ms after merger are shown.  
{\em Bottom panel}: density-weighted azimuthal and vertical average of
the energy-averaged neutrino optical depth for each species of neutrino,
computed 40~ms after merger.}
\label{fig:thickness}
\end{figure}

As can be seen from Fig.~\ref{fig:globalevolution},
as the accretion stream collides with itself, shocks heat
the gas for roughly one millisecond, until the density-averaged
entropy settles at $\sym 8\,k_B\,\rm baryon^{-1}$. 
A hot accretion disk forms in the vicinity of the black hole.  In
Fig.~\ref{fig:disksnapshot}, we show density snapshots
of the disk at a representative time $\sym 30$~ms later,
and in Fig.~\ref{fig:thickness}, we show
azimuthally averaged equatorial density as
a function of circumferential radius.  The maximum density remains at a fairly
steady level of $\sym 10^{12}\rm\,g\,cm^{-3}$.  As shown in the bottom
panel of Fig.~\ref{fig:thickness}, the densities and temperatures
are sufficient to render the interior of the torus optically thick
to all species of neutrinos, with optical depths (averaged over
neutrino energy) of order 10.

The disk is initially extremely distorted and nonaxisymmetric.  For
a completely stable disk, one would expect the inner regions, where
the dynamical timescale is shortest, to settle to a stationary
axisymmetric state before the outer regions---as was seen, for example,
in a recent BHNS $\Gamma$-law EOS merger carried out with the same
code~\citep{Lovelace:2013vma}.  (Unfortunately, we have not performed
any $\Gamma$-law EOS merger simulations with similar mass ratio and
black hole spin to the present case, so no proper comparison can be
made.)  We see instead that it is only the
middle disk that settles to an approximately axisymmetric state after
about 30\,ms. 
In the inner disk, very close to the black hole, clear, order unity
deviations from axisymmetry in the form of trailing one-armed spirals
persist throughout the evolution (see the bottom panel in Fig.
\ref{fig:disksnapshot}).  These perturbations appear at the
disk's inner edge, rotate at roughly the rate of the local fluid
($\sym\rm ms$ periods), expand outward and dissipate, and then reform
many times during the disk evolution.  Such behavior suggests that
a fluid instability may be preventing the disk from promptly settling.

\subsection{Disk equilibrium}

\begin{figure}
\includegraphics[width=8.2cm]{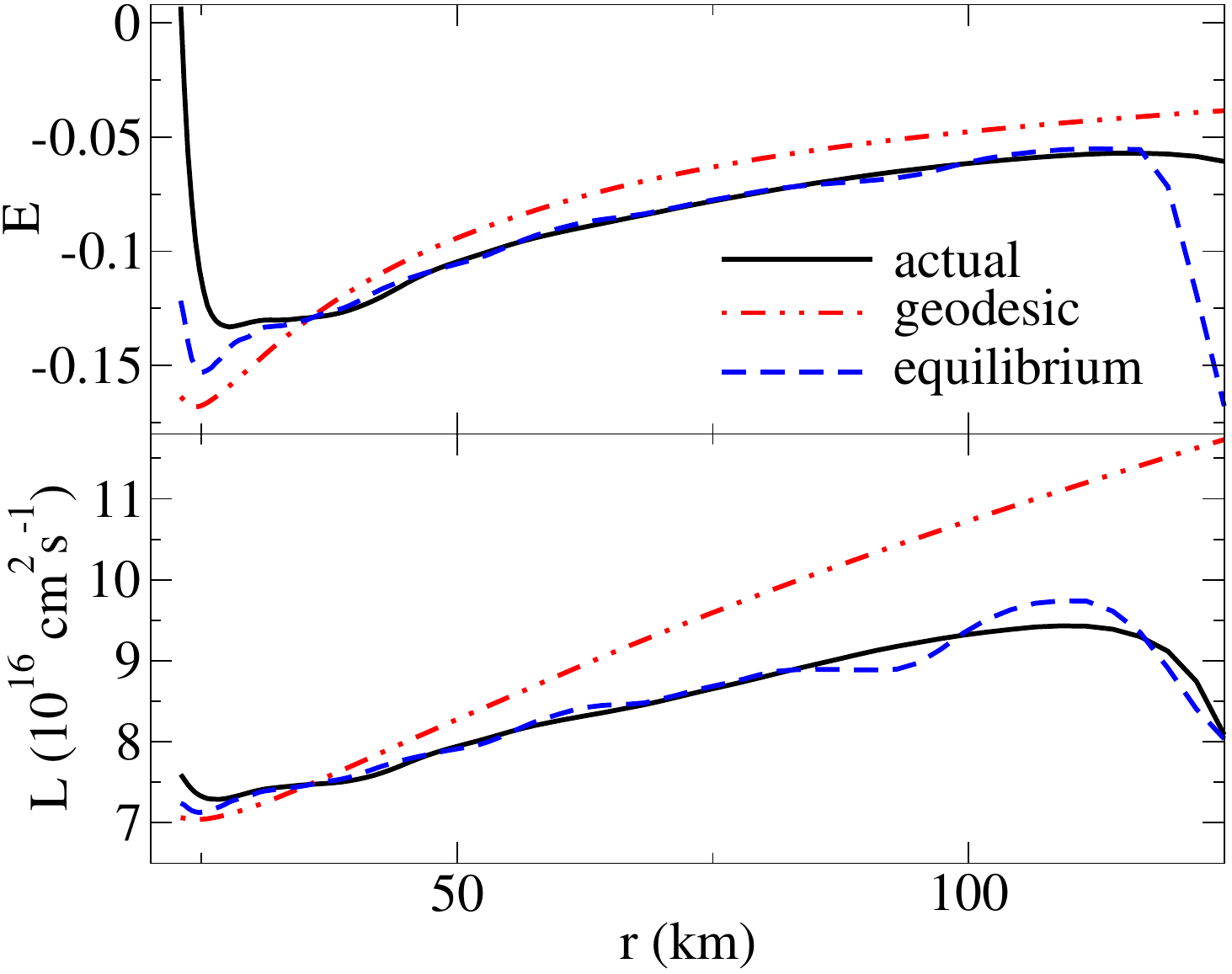}
\caption{
Energy and angular momentum of the disk 40~ms after merger.
{\em Top panel}: azimuthally-averaged equatorial orbital energy
$E\equiv-u_t-1$.
{\em Bottom panel}: azimuthally-averaged angular momentum
$L=-u_{\phi}/u_{t}$.
In addition to $E$ and $L$ for the
actual flow, we include values for circular orbit geodesic motion and
circular orbit equilibrium (i.e. including pressure) motion.
}
\label{fig:equilibrium}
\end{figure}

Further insight into the structure of the disk comes from the profiles
of the specific orbital energy ($E\equiv-u_t-1$) and angular momentum
($L=-u_{\phi}/u_{t}$) shown in Fig.~\ref{fig:equilibrium}.  Each panel
includes three curves.  One is the actual $E$ or $L$ profile, measured
on the equator and averaged over azimuthal angle.  Another is the
`Keplerian' $E$ and $L$, the values that would be found for geodesic
equatorial circular orbit given the evolved spacetime metric of the
system.  The geodesic circular orbits are those for which
$\nabla_{\vec{u}}\vec{u}=0$.  These curves thus include the effects of
the disk's self-gravity (at least, as it was at the beginning of the
Cowling evolution) but disregard pressure forces.  We see that the angular
momentum profile, for instance, is much shallower than would be found
for an equilibrium thin (geodesic) disk.  Finally, we include
`equilibrium' curves, the $E$ and $L$ for equilibrium circular
orbit given the existing metric, density, and pressure.  These
are orbits for which $\nabla_{\vec{u}}\vec{u}=-\vec{\nabla}P/(\rho h)$. 
In regions
where deviations from equilibrium are nonlinear, one must take care
in identifying these curves with the true equilibrium about which the
disk is perturbed, since $m=0$ perturbations in the pressure will
feed back into the equilibrium condition.  Given this qualification,
we see that the highest-density regions do appear to be in equilibrium
in an angle-averaged sense.  The outer disk has sub-equilibrium
angular momenta, so the gas remains in eccentric orbit.  Interestingly,
the energy and angular momentum increase dramatically at the inner
edge of the disk.  This feature is not present in the geodesic curves,
but it is in the equilibrium curves, indicating that it is a consequence
of a sharp pressure gradient near the inner edge of the disk.  Below,
we will consider the effect of these gradients on the expected stability
of the disk.  The high
energies in the inner disk would make it easy to generate an outflow there. 
We do indeed observe some mass ejection from the inner disk, but at densities
too low to be reliably modeled by our code.

\begin{figure}
\includegraphics[width=8.2cm]{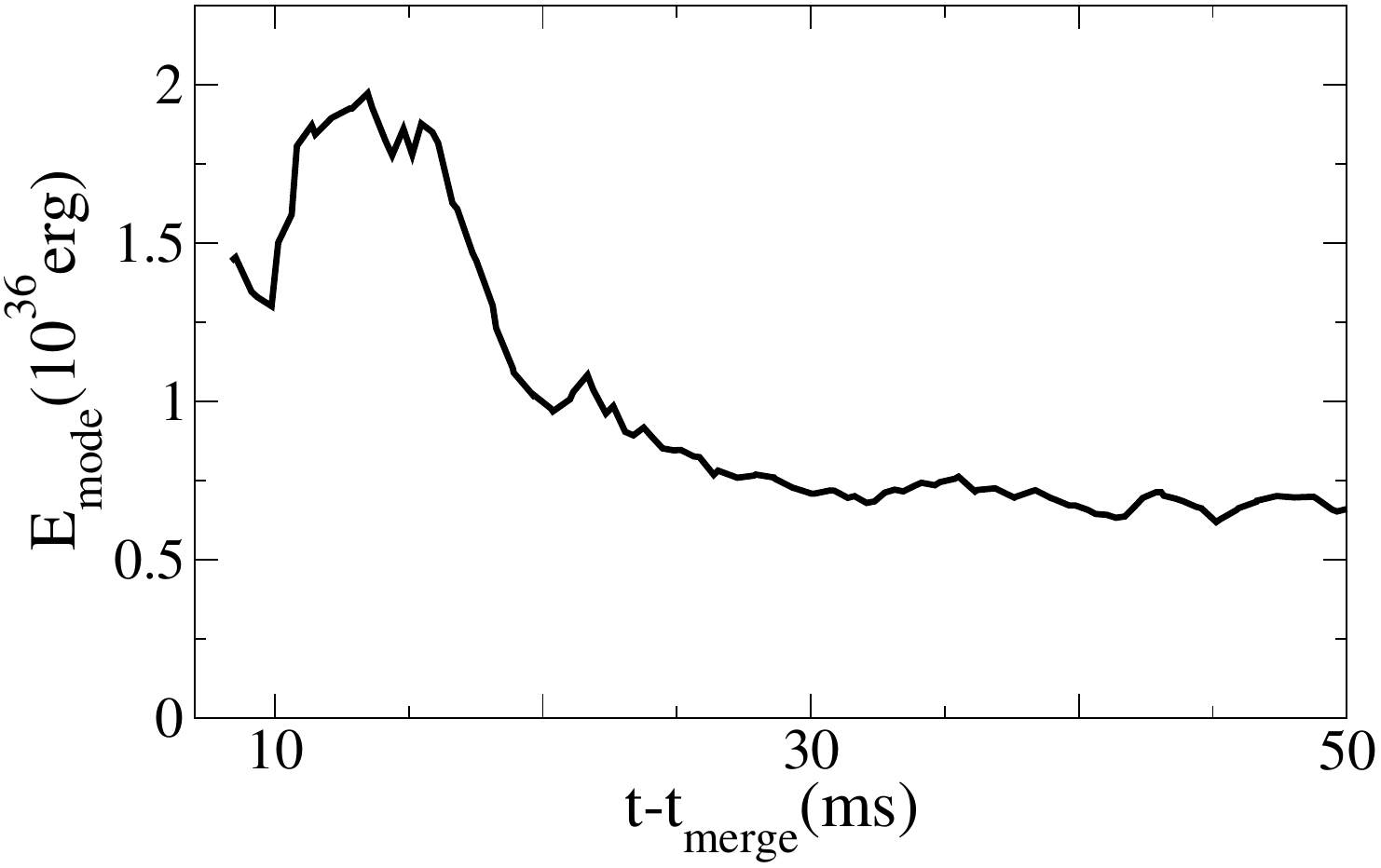}
\caption{The deviation of the accretion disk from
equilibrium, as measured by the difference between the
actual and equilibrium orbital energy.}
\label{fig:mode}
\end{figure}

The difference between the actual and equilibrium energy curves provides a
measure of the mode energy, again, to the extent that equilibrium curves
can reliably be computed from an incompletely settled pressure profile: 
$E_{\rm mode}\equiv \int \rho_*(E-E_{\rm equilibrium})d^3x$.  As shown in
Fig.~\ref{fig:mode}, 
the mode energy
is positive but decreases during the early, rapid accretion, phase.  Then
it settles and even shows episodes of growth, which may indicate that
instabilities are stimulating these modes.

\begin{figure}
\includegraphics[width=8.2cm]{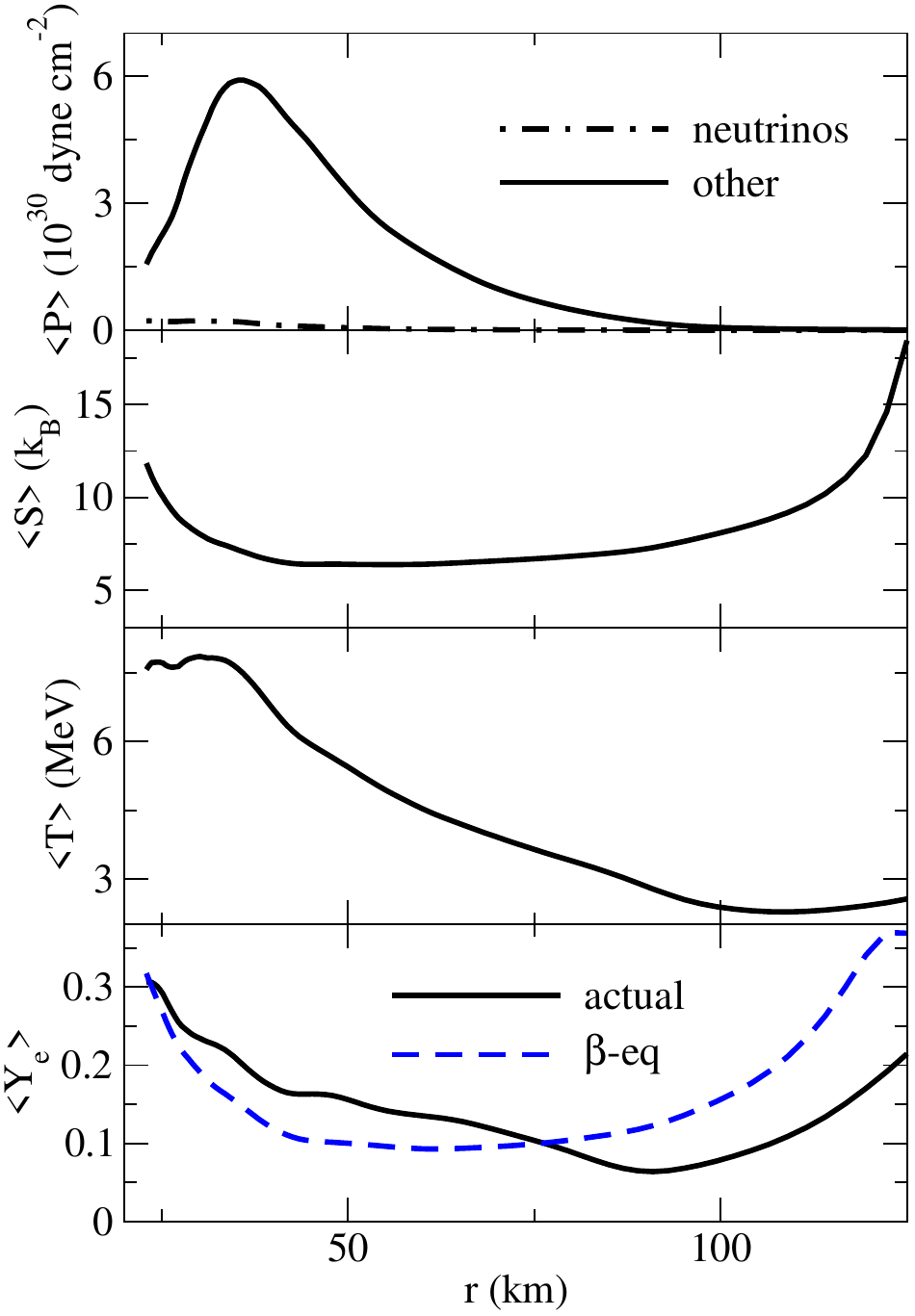}
\caption{
Profiles of the azimuthally- and vertically-averaged pressure, entropy, temperature, and
electron fraction 40~ms after merger. For the pressure, we plot separately
the contributions from the neutrinos and everything else (nucleons,
electrons, photons).  For $\langle Y_e \rangle$, we show both the actual value and the
equilibrium value (at which $R_{\nu}=R_{\nu_e}-R_{\overline{\nu}_e}=0$ for the
given density and temperature profile).
}
\label{fig:profiles}
\end{figure}

\subsection{Disk stability}

In Fig.~\ref{fig:profiles}, we plot the density-averaged pressure $P$,
entropy $S$, temperature $T$, and electron fraction $Y_e$ as functions
of circumferential radius---reducing from three dimensions by averaging
these functions vertically and azimuthally.
The entropy profile is fairly flat in
the bulk of the disk, with the exception of a steep negative gradient
($dS/dr<0$) 
in the hot inner region.  Since in this region pressure increases with
radius, this entropy gradient is actually a stabilizing force.  The $Y_e$
gradient also affects the
buoyancy,
but for these profiles,
its effect is much smaller than that of the entropy.  The relativistic
Solberg-H{\o}iland critera for convective stability of an axisymmetric
equilibrium fluid are given, for a coordinate basis in which
$g_{tr}=g_{r\phi}=0$, by \citep{1975ApJ...197..745S}
\begin{eqnarray}
\label{SH1}
\vec{\gamma}\cdot\vec{\nabla} L + (\rho h)^{-2}(\partial\mathcal{U}/\partial S)_P
\vec{\nabla} P\cdot\vec{\nabla} S &\ge& 0\ , \\
\label{SH2}
(\partial\mathcal{U}/\partial S)_P(\vec{\gamma}\times\vec{\nabla}P)\cdot
(\vec{\nabla}S\times\vec{\nabla}L) &\ge& 0\ ,
\end{eqnarray}
where $\mathcal{U}=\rho+\rho\epsilon$ is the total energy density and
\begin{equation}
\vec{\gamma} = (u^0u_0)^2[(1-v^2)(g_{\phi\phi})^{-1}u_0{}^2\vec{\nabla}L
-\vec{\nabla}\Omega]\ .
\end{equation}
Equations \ref{SH1} and \ref{SH2} correspond to the familiar Newtonian
radial and vertical stability conditions, respectively. 
The first term in Eq.~\ref{SH1} dominates everywhere inside
of $r\approx 120$\,km, so $dL/dr$ determines the stability to radial convection
through most of the disk. 
The condition for stability is met except in the inner edge ($r<27$\,km)
of the disk.  This inner region should be unstable.  The unstable
gradient persists, and it is presumably connected to the inability
of the inner disk to settle to stationary axisymmetry.  The outer
disk also has an unstable angular momentum gradient ($dL/dr<0$), but
it is stabilized by a strong positive entropy gradient. 
The entropy does not change much with height, but we do
identify a negative gradient in a region near the equator, leading to
an instability according to Eq.~\ref{SH2} there.  Buoyant forces would be
expected
to correct this gradient on timescales $(-N_{\rm BV,z}^2)^{-1/2}\sim 10$ms, but
apparently other processes (see Section~\ref{sec:neutrinodisk} on processes
that affect entropy
evolution) are sufficient to keep the entropy profile roughly fixed.

In addition to local instability, the disk could be susceptible to global
instabilities involving amplification of the unstable mode across its corotation
radius, located at $r \approx 27\rm\,km$, given the mode's pattern
speed (e.g \citealt{1985MNRAS.213..799P}).

The temperature is high ($\sym 8$~MeV) in the inner disk, and in these
regions, the contribution of trapped neutrinos to the total pressure is as
high as 12\%.  In the bulk of the disk, the neutrino pressure is
negligible.

\section{Neutrino-driven disk evolution}
\label{sec:neutrinodisk}

\subsection{Disk neutrino luminosity}

\begin{figure}
\includegraphics[width=8.2cm]{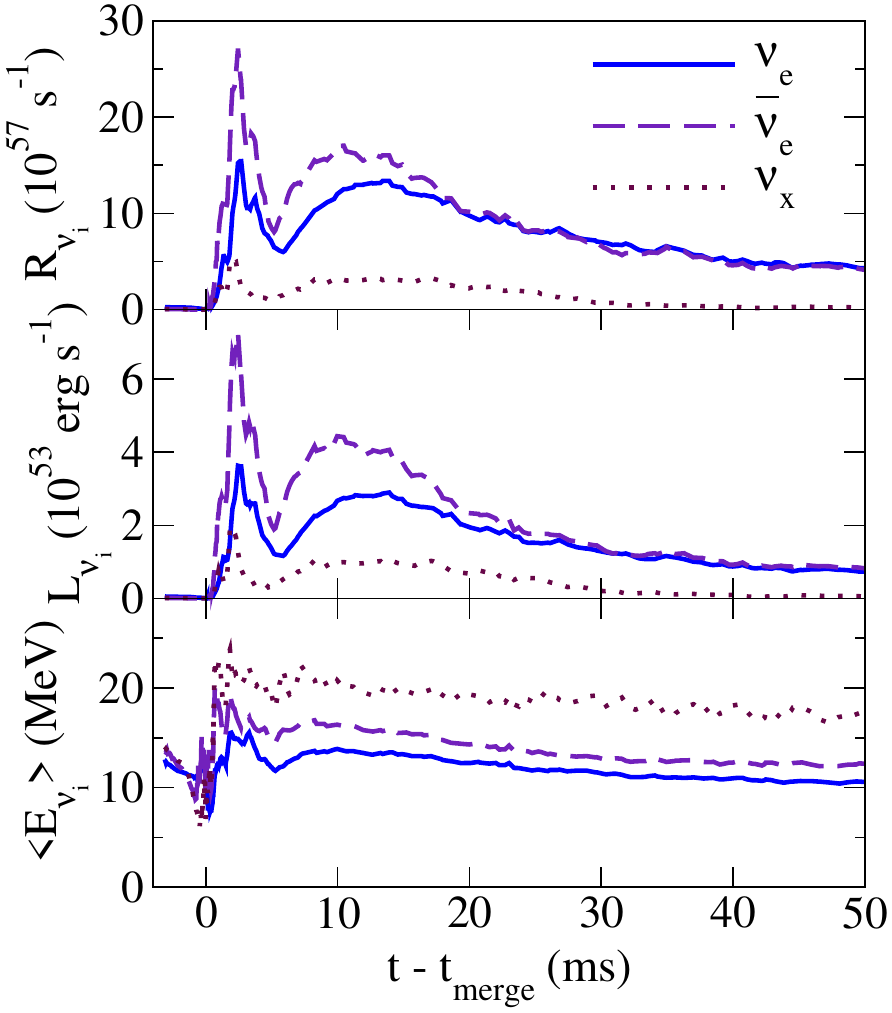}
\caption{
Neutrino emission characteristics by species.
{\em Top panel}: lepton number luminosity, $R_{\nu_i}$
(in units of $10^{57}\rm\,s^{-1}$).
{\em Middle panel}: energy luminosity, $L_{\nu_i}$
(in units of $10^{53}\rm\,erg\,s^{-1}$).
The total luminosity is given by $L_{\nu_e}+L_{\bar{\nu_e}}+4L_{\nu_x}$. 
$L_{\nu_x}$ is, thus, the luminosity of each individual $\mu$ and $\tau$
species.
{\em Bottom panel}: average neutrino energy $\langle E_{\nu_i} \rangle$
(in unit of MeV).
All plots are taken from the L3 simulation, and are calculated for
an observer at $r \rightarrow \infty$.
(Note, the bottom panel was affected by an error in our
calculation, reported in Sec.~\ref{sec:erratum}, and corrected in
Fig.~\ref{fig:fig12_correction}.)
}
\label{fig:neutrinos_by_species}
\end{figure}

Fig.~\ref{fig:neutrinos_by_species} breaks down the energy and number
emission by neutrino species.  Initially, $\overline{\nu}_e$
dominates over $\nu_e$ emission, as would be expected for a neutron-rich,
proton-poor gas.  As emission continues, protons become more numerous, and
the neutrinosphere cools, so that positrons become less common than electrons. 
(The electrons are mildly degenerate, $\mu_e/k_BT\approx 1$, so the relative
ratio of electron to positron density is quite sensitive to temperature.)
Thus, $R_{\nu_e}$ and $R_{\overline{\nu}_e}$ become closer.  At 20~ms after merger,
these emission rates are sufficiently balanced that $Y_e$ thereafter
evolves on the slightly longer timescale on
which the disk itself is changing.  In Fig.~\ref{fig:profiles}, we 
compare the actual $Y_e$ to the equilibrium $Y_e$ profile, i.e. to the
distribution $Y_e=Y_e{}^{\rm eq}$ for the given $\rho$ and $T$, that would yield
$R_{\nu}=R_{\nu_e}-R_{\overline{\nu}_e}=0$ everywhere in our leakage scheme. 
At early times,
$Y_e < Y_e{}^{\rm eq}$ for most of the matter and $Y_e$ grows.  At late times
(as in the figure), the inner disk is nearly in equilibrium, but
$Y_e > Y_e{}^{\rm eq}$ in the rest of the disk, so the average $Y_e$ decreases
slowly.  The low-density outer region radiates, and thus responds to emission
imbalances, much more slowly.  It is still in its initial $Y_e$ growth phase.
The average energy per neutrino, given by $L_{\nu_i}/R_{\nu_i}$, averaged in
time over the disk evolution, is about 12, 15, and 19~MeV for
$\nu_e$, $\overline{\nu}_e$, and $\nu_x$ neutrinos respectively;
the average neutrino energies are not constant, but decrease at a rate of 1~MeV per 10~ms.
(Note, these energy estimates were affected by an error in our calculation. They are
corrected in Sec.~\ref{sec:erratum}.)
When one adds together all four species of $\nu_x$ neutrinos, there are still fewer
of them emitted than $\nu_e$ or $\overline{\nu}_e$ neutrinos, but their
average energy is sufficiently higher ($\nu_x$ are emitted from the hotter regions
in the disk interior) that their combined luminosity is slightly larger than
$L_{\nu_e}$ and $L_{\overline{\nu}_e}$. The fact that $L_{\nu_x}$ is roughly a
quarter of $L_{\nu_e}$ may seem surprising, given that the charged current emissivity
in the luminous part of the disk is two orders of magnitude higher than the thermal
emissivity.  However, charged current processes also dominate the opacity, so that the
opacity of the brightest part of the disk is smaller for muon and tau neutrinos
($\tau_{\nu_x}\lesssim 1$) than for electron (anti)neutrinos ($\tau_{\nu_e}\sim 2-5$).

\subsection{Neutrino cooling effects}

\begin{figure}
\includegraphics[width=8.2cm]{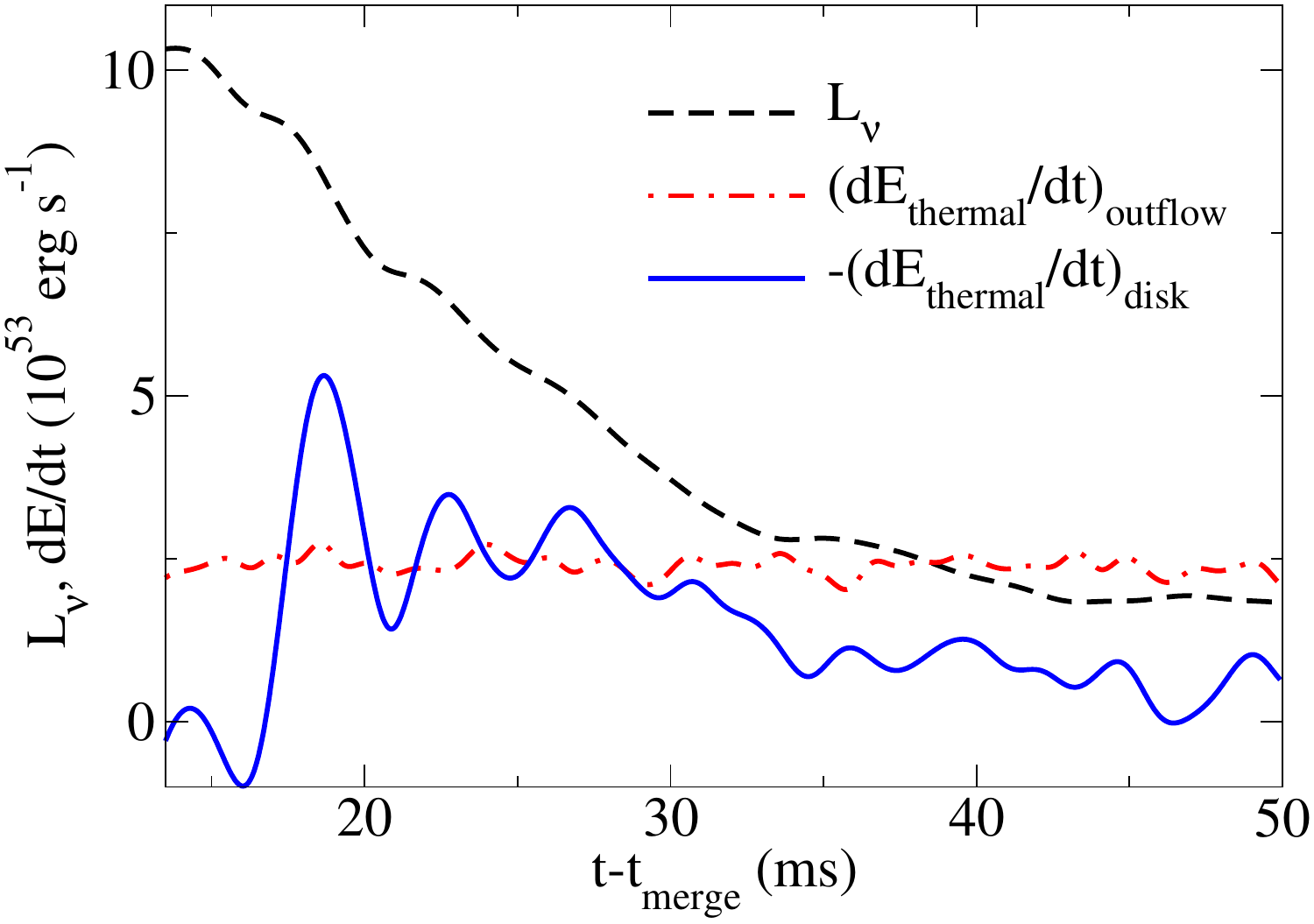}
\caption{
The thermal evolution of the disk during its cooling phase. 
$L_{\nu}$ is the neutrino luminosity, $(dE_{\rm thermal}/dt)_{\rm outflow}$
the heat loss due to flows out of the inner and outer boundaries,
and $(dE_{\rm thermal}/dt)_{\rm disk}$ the numerical derivative of the
total thermal energy in the disk.  (The total internal energy
is almost exactly $1.15E_{\rm thermal}$ throughout the Cowling evolution.) 
Numerical derivatives and
outflow measures are noisy, so the curves have been smoothed in
time by convolving with a Gaussian of width 1~ms.
}
\label{fig:cooling}
\end{figure}

From Fig.~\ref{fig:convergence}, we see that the late-time
neutrino luminosity is around $2\times 10^{53}\rm\,erg\,s^{-1}$
and continuing to drop.  During the early disk phase,
$\dot{M} \approx 1\,M_{\odot}\rm\,s^{-1}$, so the
accretion efficiency is $\eta=L_{\nu}/\dot{M}c^2 \approx 0.1$. 
The disk remains very nondegenerate throughout the simulation, with
the thermal component of the internal energy remaining nearly
a constant 85\% of the total internal energy throughout.
The actual values of the thermal and internal energies
decrease at a rate only about half $L_{\nu}$ (see Fig.~\ref{fig:cooling}).
The rate of loss of thermal energy due to flow out of the grid,
comprised at early times primarily of accretion into the black hole
and at late times primarily of disk outflow, is also comparable
to $L_{\nu}$.  Thus, these energy sinks are being countered
by heat sources sufficient to slow the cooling rate by about a
factor of 3.  Two physical sources of heating in parts of the disk are
adiabatic compression (primarily the observed flattening of the disk
toward the equator as it cools) and shock heating, the latter presumably
occuring in the nonlinearly perturbed regions of the disk.  Adiabatic
flattening alone is insufficient, since we observe that the rate of
total entropy decrease is also much lower than the radiative entropy
loss rate.  Shock dissipation is the more plausible heat source, since
losses in spiral mode energy are of the needed magnitude.
A third physical energy source, recombination of nuclei, does not occur
during the phase of evolution simulated;
the composition of the disk remains $>99$\% free nucleons. 
A nonphysical source of heating would be numerical viscosity. 
This can only be distinguished from true shock heating by
comparing heating rates at different resolutions.  We find that
L2 and L3 have very similar thermal histories, while L1
cools 30\% more slowly, largely due to a difference in the outflows,
although effects of stronger numerical energy dissipation might
also contribute.  The average specific
entropy of the disk provides a clearer sense of whether the remaining
matter is heating or cooling.  Its evolution is included in
Fig.~\ref{fig:globalevolution}.  We see that, after the initial
strong shock heating at disk formation, the entropy begins a slow
decrease.  The fact that different resolutions show good agreement
on the entropy evolution reassures us that numerical dissipation
is not dominating the thermal evolution of the disk.

\section{Comparison with evolution without neutrino cooling}
\label{sec:comparison}

At least for the very massive, luminous post-merger system studied
here, neutrino cooling and composition changes have a strong effect
on the disk structure, even in the early tens of milliseconds.  We
estimate these effects by comparing simulations carried out with (L1) and
without (L1no$\nu$) the neutrino leakage source terms added to the fluid
equations.

The disk evolved without neutrino cooling develops a significantly
higher, and continually growing, entropy.  This leads to a larger,
more diffuse accretion disk; the maximum density is a factor of ten lower
than for the neutrino-cooled disk, and the density profile lacks the
sharp peak in the inner region seen in Fig.~\ref{fig:thickness}. 
The larger extent of the non-cooled disk causes more matter to escape
through the outer boundary.  During the first 10~ms after merger, the
accretion rate into the black hole for cooled and non-cooled disks is
comparable, and so the L1no$\nu$ disk comes to have a baryonic mass lower by
about 40\% than the L1 disk.  Again, this is a consequence rather than
the cause of the lower density.  At later times, the accretion rate
of the L1no$\nu$ disk drops to very low values, but outflow continues. 

Because of its lower density, the
non-cooled disk actually has a lower average temperature than the
cooled disk (4~MeV vs.\ 5~MeV), despite its higher entropy, during the
first 45~ms after merger.  After this, the temperature of the L1 disk
drops below that of the L1no$\nu$ disk because of neutrino cooling.  As
expected, neutrino emission strongly affects the evolution of $Y_e$. 
For the L1no$\nu$ disk, the electron fraction can only change because of
the accreted or outflowing matter having $Y_e$ that differs from the
average, and this turns out to cause only small changes to the
average (see Fig.~\ref{fig:globalevolution}).

The L1no$\nu$ disk shows the same persistent perturbations from axisymmetry
that are seen in the neutrino-cooled disks.  High-energy, non-equilibrium
flow in the inner disk is still seen.  This confirms that cooling is not
the main driver
of perturbations in the cooled disks.  We find that the geodesic and equilibrium
angular momentum curves show greater deviation for the L1no$\nu$ disk---
the geodesic curves are essentially the same, but $dL/dr$ is about 25\%
shallower for the equilibrium curve in the bulk of the non-cooled disk---
indicating that pressure support makes a stronger contribution to the
equilibrium of this disk.  The entropy profiles have the same general
shape, and the expected stability properties are similar.

\section{Conclusions and Future Work}
\label{sec:conclusion}

In this paper, we present simulations of a black hole--neutron star
binary merger event using a hot nuclear theory-based equation of state with
neutrino cooling.  This allows us to address features of the
merger---composition changes, neutrino energy extraction---that are inaccessible
to models with any less degree of realism.  While important simulations
with hot EOS and neutrino leakage have been done in the past, we perform
them for the first time in full general relativity.  We are thus able
to include distinctly relativistic features, such as a high black hole
spin.

For this high-spin, moderate-mass system, we find that a significant amount
of matter evades prompt capture by the black hole.  Roughly 0.08~$M_{\odot}$
of low-$Y_e$ tidal tail material appears to be dynamically ejected from the
system at mildly relativistic speeds.  This material will contribute to
r-process enrichment of the interstellar medium, and may produce quasi-isotropic
electromagnetic signals in the form of an optical or infrared kilonova and a radio afterglow,
although our simulations do not model such processes.
Because the opacity properties of the exotic nuclei formed in the tail are so poorly
constrained (although see \citealt{Kasen:2013xka} for progress on this problem),
quantitative measures of kilonova detectability are subject to large uncertainties.
\cite{metzger:11} do, however, define a quantitative figure of merit for
the detectability of the radio transient,
\begin{equation}
  \label{eq:FOM_rad}
  {\rm FOM}_{\rm rad} = \left(\frac{E_{\rm ej}}{10^{50}\rm\,erg}\right)
  \left(\frac{n}{1\rm\,cm^{-3}}\right)^{7/8}
  \left(\frac{v}{c}\right)^{11/4},
\end{equation}
where $n$ is the number density of the circumbinary medium.
For the ejecta measured in this simulation, and assuming an optimistic
circumburst density of $0.1\rm\,cm^{-3}$ \citep{Berger:2005rv,Soderberg:2006bn},
we roughly estimate ${\rm FOM}_{\rm rad} \approx 0.15$,
which approaches the detection threshold of 0.2
given for the Expanded Very Large Array
(EVLA)\footnote{\url{http://www.aoc.nrao.edu/evla/}} at 1~GHz, for
events within a range of 200~Mpc.

Of the matter that remains bound, 0.3~$M_{\odot}$ forms a hot,
neutrino-optically-thick accretion disk.  We find that neutrino cooling has a significant
effect on the disk structure; uncooled disks are significantly more extended. 
Nevertheless, the disk remains hot and thick, with markedly non-Keplerian
orbital profiles.  The innermost disk configuration appears to be unstable,
and order
unity deviations from equilibrium and axisymmetry, in the form of
millisecond-period spiral features, persist for tens of milliseconds after
merger.  The neutrino luminosity after merger is initially very high
($10^{54}\rm\,erg\,s^{-1}$), but disk depletion due to accretion and outflow,
and energy depletion due to radiation cause the disk to quickly dim.  Its luminosity
is able to remain above $2 \times 10^{53}\rm\,erg\,s^{-1}$ for about 50~ms.

Energy deposition efficiencies of 0.2\%--0.5\% have been seen in prior studies
of $\nu\overline{\nu}$ annihilation near similar accretions disks
\citep{1997A&A...319..122R,Birkl2007,dessart2009,Harikae:2010yt}.
In this particular optimistic case (optimistic because of the energetic, long-lived disk),
the $\nu\overline{\nu}$ annihilation mechanism may be able to
power a fireball with an energy within the range of observed SGRBs.
To examine the $\nu\overline{\nu}$ efficiency in this scenario,
a quantitative analysis is needed, accounting for the spatial and spectral distribution
of neutrinos, and the relativistic effects due to the spinning black hole.
This could be done, for example, by a ray-tracing solution of the Boltzmann equation.

Additionally, the drop in luminosity at late times
is most likely an artifact of the absence in this simulation of physical
processes expected to dominate the long-term evolution of the disk.  Most
obviously, accretion disks like this one will be subject to the
magnetorotational instability~\citep{BalbusHawley1991}.  The MRI can drive vigorous turbulence
which, through viscosity or reconnection at small scales, can reheat
the disk.  Independent of its possible effects on the thermal power
available, magnetic forces could make possible Poynting flux-dominated
outflows, tapping either the rotational energy of the disk or the spin
energy of the black hole.

Although it is undoubtedly interesting, the system studied here is only
one case of BHNS merger, and perhaps not a typical one.  It will be
important to carry out simulations of more systems, covering a range of
neutron star masses, black hole masses, and black hole spins.  We believe
our code captures enough of the relevant physics to be suitable for
such explorations; it should certainly improve upon the
polytropic neutron star models we and others have previously used for such surveys.
Simulations with this level of realism could also be used to study the effects of
the different choices of hot EOS used to simulate a merger.
Several hot nuclear EOSs have been made
publicly available and are suitable for codes like ours
(\citealt{Lattimer:1991nc};
\citealt{Shen:1998gq};
\citealt{Shen:2011kr};
\citealt{Hempel:2011mk};
\citealt{Steiner:2012rk}).

We hope to make several improvements to future simulations. Much could be gained
for nucleosynthesis estimates by tracking the tidal tail out to farther distances
and with higher accuracy. To simultaneously resolve outgoing ejecta and the
incipient accretion disk is a challenging task for any grid-based code. 
The solution will likely involve some combination of increased computational resources
and improved modeling techniques, such as adaptive meshing or Lagrangian
(SPH) evolutions of the outflow.  We are also working to incorporate
missing crucial pieces of physics.  Magnetic effects require
magnetohydrodynamic simulations, while the effects of neutrino heating
will only be accessible to our models when we replace the leakage scheme
with some form of genuine radiation transport.

\section*{Acknowledgements}

We acknowledge helpful discussions with J.~Lattimer, L.~Roberts, and
S.~Teukolsky.  This research is supported in part by NASA ATP Grant
No. NNX11AC37G, by the National Science Foundation under grand
Nos. PHY-1068243, PHY-106881, PHY-1151197, and AST-1205732, by the
Alfred P. Sloan Foundation, and by the Sherman Fairchild
Foundation. The computations were performed at Caltech's Center for
Advanced Computing Research on the cluster `Zwicky' funded through NSF
grant no.\ PHY-0960291 and the Sherman Fairchild
Foundation. Furthermore, computations were performed on the General
Purpose Cluster supercomputer at the SciNet HPC Consortium. SciNet is
funded by: the Canada Foundation for Innovation under the auspices of
Compute Canada; the Government of Ontario; Ontario Research Fund--Research
Excellence; and the University of Toronto \citep{SciNet}.

\section{Erratum (Fall 2015)}
\label{sec:erratum}

\subsection{Error in $\protect\Rnui$}

We recently discovered an error in the leakage calculations which incorrectly
used \code{log10} instead of \code{log} in the analytic form of the
zeroth-order Fermi Integral.
The integral was used to compute number loss rates in the diffusive
regime, as in Eqn.~A34 in Rosswog and Liebend{\"o}rfer, MNRAS 342, 673 (2003).
The effect of the error was to artificially suppress neutrino number emission
rates, $R_{\nu_i}$, in the optically thick regions of the simulations by a factor
of approximately $\log_{10}{e}\sim0.5$.
It had no direct effect on the energy emission rates, $Q_{\nu_i}$.
Below we estimate the effect of our error on the average electron fraction,
$\langle\Ye\rangle$, and average neutrino energies, $\avgE$.

\subsection{Effect on $\langle\protect\Ye\rangle$} 
From comparisons of a disk simulation (different from this one, but also
experiencing significant diffusive neutrino losses) computed with and without
the error, we conclude that $\langle\Ye\rangle$ was not significantly affected.
This is expected. For most of its life the disk is in a quasi-equilibrium state
with respect to the charged-current reactions. In this regime the timescale for
a local change in $\Ye$ (set by $R_{\nu_e}-R_{\bar{\nu}_e}$) is much shorter than
for changes in density and temperature.
A small difference in the rate of change of $\Ye$ doesn't effect
this hierarchy of timescales, and the disk's $\langle\Ye\rangle$ is simply set
by its density and temperature.
For further description of this comparison see Foucart et al.\ Phys.\ Rev.\ D
93, 044019, 2016, Sec.\ II.C.

In addition, the neutrino-driven evolution of the dynamical outflows was
dominated by free-streaming neutrino losses, so that the effect of the error was
negligible also on our estimate of the ejecta $\langle Y_e \rangle$.

\subsection{Effect on $\protect\avgE$} 
The only significant effect of this error on our findings was that
our estimates of the average neutrino energies were artificially high.
This is because the total number-averaged energies were calculated from
$\langle E_{\nu_i} \rangle \equiv \int d^3x\, Q_{\nu_i}/\int d^3x\, R_{\nu_i}$,
and the error suppressed $R_{\nu_i}$ by a factor of 2 (at most)
while leaving $Q_{\nu_i}$ unchanged.

Here we recalculate the neutrino energies at several times late in the simulation.
Fig.~\ref{fig:fig12_correction} corrects the bottom panel of Fig.~12 from the
original paper.
The sentence interpreting that figure on p.~13 should now read:
``The average energy per neutrino, given by $L_{\nu_i}/R_{\nu_i}$,
averaged in time over the disk evolution is about
7, 9, and 14~MeV
for $\nu_e$, $\bar{\nu}_e$, and $\nu_x$ neutrinos respectively;
the average neutrino energies are not constant, but decrease
for $\nu_e$ and $\bar{\nu}_e$ at a rate of about 0.5~MeV per 10~ms.
The brief increase in $\langle E_{\nu_x} \rangle$ is a transient effect
of the transition of that species' emission from an epoch dominated by diffusion
to one dominated by free-streaming.''

\subsection{Conclusion}
\begin{figure}
\vspace{0.0cm}
\includegraphics[width=\linewidth]{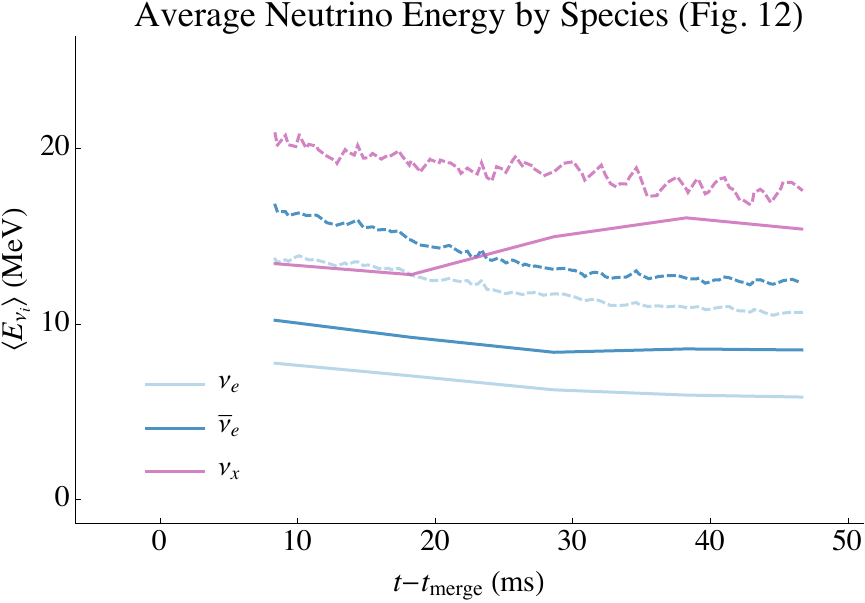}
\caption{
Average neutrino energies for each species. The dashed lines are reproduced
from the last 40~ms of the bottom panel of Fig.~12 of the original paper;
the solid lines are the corrected estimates.
Note, as the disk cools the boundary separating the regions in which neutrinos
diffuse from those in which they freely escape encloses less and less of the
matter, and the effect of the error on the average neutrino energies becomes less
and less significant.
This is especially pronounced for $\nu_x$, for which the erroneous energy estimates
almost agree with the corrected estimates by the end of the simulation.
}
\label{fig:fig12_correction}
\vspace{0.5cm}
\end{figure}

This code error led us to report average neutrino energies that were erroneously
high for a leakage model. See Fig.~\ref{fig:fig12_correction} for a correction.
It had an insignificant effect on the hydrodynamics and composition evolution.

\bibliography{References/References}

\end{document}